\documentclass[12pt]{article}

\usepackage[normalem]{ulem} %

\usepackage{amsthm,amsmath,mathrsfs,amsfonts,amssymb,mathtools,nicefrac}
\usepackage{bbm}
\usepackage{soul}

\usepackage{graphicx,caption,subcaption}

\usepackage{natbib}
\usepackage{color}
\definecolor{dgreen}{rgb}{0,0.5,0}
\usepackage{hyperref}
\hypersetup{
    breaklinks=true,
    colorlinks,
    citecolor=blue,
    filecolor=black,
    linkcolor=blue,
    urlcolor=black
}
\usepackage{url}

\usepackage[multiple]{footmisc} \usepackage[capposition=top]{floatrow}

\usepackage{enumerate}

\usepackage{adjustbox}
\usepackage{multirow}

\usepackage{amsmath}
\usepackage{accents}
\newlength{\dhatheight}

\newcommand{\plim}{\operatorname*{plim}}

\newcommand{\blind}{1}
\newtheorem{assumption}{Assumption}
\newtheorem{definition}{Definition}

\newtheorem{theorem}{Theorem}
\newtheorem{lemma}{Lemma}

\theoremstyle{definition}

\newtheorem{remark}{Remark}

\usepackage[margin=1.25in]{geometry}
\usepackage{setspace}
\usepackage{titlesec}
\usepackage{lmodern}

\begin{document}

\def\spacingset#1{\renewcommand{\baselinestretch}{#1}\small\normalsize} \spacingset{1}

\if1\blind
{
  \title{\vspace{-1cm}{\bf  \large \shortstack{A NEYMAN-ORTHOGONALIZATION APPROACH TO\\THE INCIDENTAL PARAMETER PROBLEM\\IN LIKELIHOOD MODELS\thanks{We are grateful to the Editor and referees, as well as Dmitry Arkhangelsky, Jin Hahn, Bo Honor\'e, Roger Moon, Whitney Newey, Andres Santos, Vira Semenova, and Vasilis Syrgkanis for comments and discussion. \newline Funded by the European Union (ERC-NETWORK-101044319) and by the French Government and the French National Research Agency under the Investissements d'Avenir program (ANR-17-EURE-0010). Views and opinions expressed are however those of the authors only and do not necessarily reflect those of the European Union or the European Research Council. Neither the European Union nor the granting authority can be held responsible for them. }}}}
  \author{\vspace{-0.2cm} St\'ephane Bonhomme\thanks{sbonhomme@uchicago.edu} \\ {\small Department of Economics, University of Chicago} \and  \vspace{-0.2cm}  Koen Jochmans\thanks{koen.jochmans@tse-fr.eu} \\  {\small Toulouse School of Economics, Universit\'e Toulouse Capitole} \and \vspace{-0.2cm}  Martin Weidner\thanks{martin.weidner@economics.ox.ac.uk} \\{\small Department of Economics and Nuffield College, University of Oxford}}
\date{ \small August  2026}
  \maketitle
} \fi

\if0\blind
{
  \bigskip
  \bigskip
  \bigskip
  \begin{center}
    {\bf {\normalsize }}
\end{center}
  \medskip
} \fi

\vspace{-0.95cm}
\begin{abstract}
\noindent A popular approach to perform inference on a target parameter in the presence of nuisance parameters is to construct estimating equations that are orthogonal to the nuisance parameters, in the sense that their expected first derivative is zero. Such first-order orthogonalization allows the estimator of the nuisance parameters to converge at a slower-than-parametric rate. It may, however, not suffice when the nuisance parameters are very imprecisely estimated. Leading examples are models for panel and network data that feature fixed effects. In this paper, we show how, in the conditional-likelihood setting, estimating equations can be constructed that are orthogonal to any chosen order $q$, in that their leading $q$ expected derivatives are zero. This yields estimators of target parameters that are unaffected by the presence of nuisance parameters to order $q$. In an empirical illustration, we apply our method to a fixed-effect model of team production. 
\end{abstract}



\newpage

\onehalfspacing

\renewcommand{\theequation}{\arabic{section}.\arabic{equation}}  \setcounter{equation}{0}

\section{Introduction}
Inference in the presence of nuisance parameters has received substantial attention. One fruitful way to proceed is to work with estimating equations that are orthogonal with respect to the nuisance parameters in the sense of \cite{Neyman1959}. Such equations underlie many of the results in semiparametric estimation (\citealp{Newey1994}) and are at the heart of recent advances on doubly-robust estimation and high-dimensional inference as discussed in, for example, \cite*{ChernozhukovChetverikovDemirerDufloHansenNeweyRobins2018}. A key finding is that Neyman-orthogonality permits the construction of asymptotically unbiased estimators that converge at the usual $n^{-\nicefrac{1}{2}}$-rate provided the nuisance parameter has a convergence rate that is faster than $n^{-\nicefrac{1}{4}}$, where $n$ is the sample size.

However, the faster-than-$n^{-\nicefrac{1}{4}}$ requirement often fails in problems where the dimension of the nuisance parameter is large relative to the sample size. Examples are panel data models with fixed effects, which are widely used in linear and nonlinear difference-in-differences settings. There, we observe $N$ units over $T$ periods of time and the model includes both common parameters and unit-specific nuisance parameters {(such as heterogeneous intercepts or slopes)}. The latter are estimated at the rate $T^{-\nicefrac{1}{2}}$. For an estimator of the former based on Neyman-orthogonalization to be successful we would therefore need that $T^{-\nicefrac{1}{2}} = o((N T)^{-\nicefrac{1}{4}})$, which translates into the requirement that $ N= o(T)$. This is usually not a realistic condition in microeconometric applications. In fact, under this requirement the standard fixed-effect estimator would permit asymptotically-valid inference. Consequently, (first-order) Neyman-orthogonalization does not solve the incidental parameter problem in panel data.\footnote{The problem is reminiscent of the poor performance of double machine-learning techniques in some settings, as recently documented by  \cite{WuthrichZhu2021} and \cite{AngristFrandsen2022}. 
A related problem where the conventional approach was formally shown to fail
is a nonlinear version of the judge-leniency design, see \citet{hahn2021problems}.
}

The issue can be even more severe in high-dimensional regressions on network data. In such settings, the convergence rate of the estimator of the nuisance parameter depends on the connectivity structure of the network (\citealp{JochmansWeidner2019}). Examples include the estimation of teacher value-added (\citealp{JacksonRockoffStaiger2014}), of the contributions of worker and firm heterogeneity to the variance of log wages and other covariance components (\citealp{AbowdKramarzMargolis1999}, \citealp{KlineSaggioSoelvsten2020}), as well as of complementarity patterns in team production (\citealp{AhmadpoodJones2019}, \citealp{Bonhomme2021}). Fixed effects in network-formation models are also poorly estimated, especially in the prevalent case where the network is sparse (see, e.g., \citealp{graham2020sparse}).

Motivated by these concerns, we focus on a higher-order generalization of Neyman-orthogonality that was proposed by \cite{MackeySyrgkanisZadik2018}. An estimating equation is Neyman-orthogonal to order $q$ when all $q$ leading derivatives with respect to the nuisance parameter have zero expectation. When $q=1$, this means that the expected Jacobian is zero, which recovers the conventional notion of Neyman-orthogonality (to order one). Working with estimating equations that are Neyman-orthogonal to order $q$, when combined with sample splitting, allows one to construct asymptotically-linear estimators when nuisance parameters are estimated at a rate faster than $n^{-\nicefrac{1}{2(q+1)}}$. As an example, in the panel data problem, this yields valid inference under the requirement that $N=o(T^{q})$. Combining orthogonalization with sample splitting (or cross-fitting) is important to achieve such an improvement, because orthogonalized estimating equations, by themselves, do not, in general, deliver estimators with improved sampling properties.

Working in the conditional-likelihood setting, we show how to construct estimating equations that are orthogonal to any chosen order. These estimating equations can be understood to be generalizations of the projected score of \cite{SmallMcLeish1989} and \cite{WatermanLindsay1996}. 
They can also be seen as higher-order influence functions, as in \citet{RobinsLiTchetgenTchetgenvanderVaart2008} and \citet{vanderVaart2014}. Our approach applies to general low-dimensional target parameters that satisfy a moment restriction. This includes functions of the nuisance parameters such as average elasticities or other average effects. The conditional-likelihood setting is crucial for our approach, since it allows us to orthogonalize a given estimating equation without introducing additional nuisance parameters. Avoiding such additional nuisance parameters is not essential to achieve orthogonality to order one, as is well known. It is, however, what enables the construction of higher-order orthogonalized estimating equations, as we explain in Subsection \ref{subsec_intuition}. We further discuss the non-likelihood case in Section \ref{sec_final}.

Our approach relates to techniques to correct for bias in panel data (\citealp{HahnNewey2004}, \citealp{DhaeneJochmans2015b,DhaeneJochmans2015a}) and to the literature on small measurement error (\citealp{Chesher1991}, \citealp{evdokimov2023simple}). However, in contrast with these approaches, we do not restrict the nuisance parameters beyond the fact that they can be estimated at a certain rate.

We illustrate the usefulness of our approach in several examples and in an empirical application to the estimation of nonlinear regressions on network data; a problem for which, at present, no alternative solutions exist.
In this setting, we estimate a constant elasticity of substitution (CES) production function from the scientific output of research collaborations. As in \cite{AhmadpoodJones2019}, the production function depends on researcher-specific fixed effects. %
Estimates of the parameters can be used to quantify the degree of complementarity among researchers within teams, and to compute the impact of counterfactual re-allocations in the spirit of earlier work by \citet{graham2014complementarity}.  

This problem is difficult because in the data that we use (taken from \citealp{DuctorFafchampsGoyalvanderLeij2014} and concerning publications in economics on EconLit), the number of collaborations per researcher is quite low. A conventional estimator is thus likely to suffer from substantial bias. Our procedure uncovers the presence of complementarity among authors in the production of research articles. In a counterfactual exercise we also find that randomly pairing researchers would lead to a decrease in the average quality of articles. Our findings are corroborated in a simulation experiment targeted to our empirical application.

\renewcommand{\theequation}{\arabic{section}.\arabic{equation}}\setcounter{equation}{0}

\section{Problem statement and motivation\label{sec_mod}}

\subsection{Setup} %
\label{subsec:Setup}

Let $Z_i=(Y_i,X_i)$ be random vectors, for $i=1,..,N$. We consider a setting where the conditional density function of $Y_i$ at $y$ given $X_i=x$, say $\ell(y\,|\,  x;\theta_0,\eta_{i0})$, is known up to the parameters $\theta_0$ and $\eta_{i0}$. Throughout, we will treat $\eta_{10},\ldots,\eta_{N0}$ as nuisance parameters, and leave the marginal density of the conditioning variable, $\ell_{X_i}(x)$, unrestricted. We are interested in estimating a parameter $\mu_0$ that is defined through the moment condition
\begin{align}
\mathbb{E}\left(
\sum_{i=1}^N u(Z_i;\theta_0,\eta_{i0} ,\mu_0)\right) = 0,
    \label{MainMoment}
\end{align}
where the expectations are over $Z_i$ under $\ell(y\,|\,  x;\theta_0,\eta_{i0}) \,\ell_{X_i}(x)$. While the function $u$ could additionally depend on $i$, for instance in settings where the dimension of $\eta_{i0}$ differs across $i$, we omit this dependence for conciseness. We assume that, for all $i=1,\ldots,N$, $Z_i$ contains $n_i$ individual observations, and denote the total number of observations as   $n=\sum_{i=1}^N n_i$. For example, in a balanced panel data setting with $N$ units and $T$ time periods, $Z_i$ is the time series of unit $i$'s observations, $n_i=T$ for all $i$, and $n=NT$.

Our setup accommodates different types of target parameters. As an example, we can set $\mu_0 = \theta_0$. In this case, using $u(z;\theta,\eta_i)$ as a shorthand for $u(z;\theta,\eta_i,\theta)$, one possibility is to use the score,
$$
u(z;\theta,\eta_i)
=
\frac{\partial \log \ell(y\,|\,  x;\theta,\eta_i)}{\partial\theta}.
$$
More generally, the moment condition \eqref{MainMoment} defines the target parameter
$$\mu_0=\mu(\theta_0,\eta_{10},\ldots,\eta_{N0},\ell_{X_1},\ldots, \ell_{X_N}),$$ which can be a function of the
parameters $\theta_0$ and $\eta_{i0}$ describing the conditional distribution of $Y_i$ given $X_i$,
of the marginal distribution of $X_i$, and (implicitly) of the sample size. For example, we may be interested in an average effect of the form
$$
\mu_0 = \frac{1}{N}\sum_{i=1}^N\int  m(x;\theta_0,\eta_{i0})\ell_{X_i}(x) \, dx,
$$
where {$m$ is a known function.} 

To illustrate the setup we will refer to two leading examples.

\paragraph{Example: Neyman-Scott model.}

Our first example is the well-known \cite{NeymanScott1948} model. Here, 
	\begin{equation}Y_{ij}=\eta_{i0} + \varepsilon_{ij},\quad  \varepsilon_{ij}\sim \mathrm{iid}~{\cal{N}}\left(0,\sigma_0^2\right),\quad i=1,\ldots,N,\quad j=1,\ldots,T,\label{eq_neyman_scott}
    \end{equation}
	and the goal is to estimate $\theta_0 = \sigma_0^2$ in the presence of the nuisance parameters $\eta_{10},\ldots,\eta_{N0}$. Define, for all $i=1,\ldots,N$,
\begin{equation}\label{eq_NS}
 u(Y_i;\sigma^2,\eta_i)=-\frac{T}{2\sigma^2}+\frac{1}{2\sigma^4}\sum_{j=1}^T(Y_{ij}-\eta_i)^2,
\end{equation}
 where $Y_i=(Y_{i1},\ldots,Y_{iT})^\top$ has dimension $n_i=T$, and the total number of observations is $n=NT$. It is well-known that the maximum-likelihood estimator of $\sigma_0^2$ is on average too small, suffering from bias $- \sigma_0^2 / T$. While in this panel data problem first-order orthogonality does not reduce the order of this bias, we demonstrate below that second-order orthogonalization fully removes it. 

 \paragraph{Example: CES production function.} Consider an environment where we observe workers producing output in $n$ teams of size 2. Moreover, let $k(j,1)$ and $k(j,2)$ denote the workers in team $j$, and write ${\cal{K}}=\{(k(j,1),k(j,2))\,:\, j=1,\ldots,n\}$ for the set of workers in all teams; note that a given worker may be part of multiple teams.
 Consider a model for team production where team output is a CES aggregate of worker inputs (as in \citealp{AhmadpoodJones2019}),
	\begin{equation}Y_{j}= \left(\frac{\eta_{k(j,1)0}^{\gamma_0}+\eta_{k(j,2)0}^{\gamma_0}}{2}\right)^{\frac{1}{\gamma_0}}\varepsilon_{j}^{\sigma_0},\quad  \log \varepsilon_{j}\,|\, {\cal{K}}\sim \mathrm{iid}~{\cal{N}}\left(0,1\right),\quad j=1,\ldots,n.\label{eq_CES_size2}\end{equation}	
In this model, one may be interested in estimating the substitution parameter $\gamma_0$ or the log error variance $\sigma_0^2$, average elasticities, or effects of counterfactual re-allocations of workers to teams, for example.\footnote{The model relies on the assumption that the network ${\cal{K}}$ of co-workers is exogenous, i.e., independent of the shocks $\varepsilon_{j}$. Relaxing exogeneity through a parametric model of team formation is conceptually feasible within our likelihood approach, but we do not consider this extension here.}

To analyze this example we consider $N\leq n$ subsets of teams $j$, of size $n_i$ each. Let $Y_i$ denote the vector of team outcomes in subset $i$, let ${\cal{K}}_i$ denote the set of indicators for workers belonging to those teams, and let $\eta_i$ be the collection of all fixed effects of those workers. Finally, let $\theta=(\gamma,\sigma^2)^\top$. The scores with respect to $\gamma$ and $\sigma^2$ take the form $u(Y_i,{\cal{K}}_i;\theta,\eta_i)$. In contrast to our previous example, the theoretical literature on network models such as (\ref{eq_CES_size2}) is scarce, and to our knowledge no approach has as yet been developed for achieving bias reduction in such a setting. 

In this example, a worker's fixed effect may appear in the nuisance parameter $\eta_i$ across multiple observations. While such ``overlapping fixed effects'' typically complicate the analysis and correction of incidental parameter bias, our orthogonalization and estimation methods straightforwardly accommodate this structure.

\subsection{The role of first-order orthogonality and its limitations\label{subsec_22}}

In the remainder of this section, we motivate our approach in a setting where one wishes to estimate $\mu_0=\theta_0$ based on a random sample $Z_1,\ldots, Z_n$, taking $u$ to be a univariate function, and $\eta_0$ {and $\theta_0$} to be scalar. Hence $N=1$, and $n_1=n$ is the total number of observations. 

If $\mathbb{E}(\sum_{j=1}^nu(Z_j;\theta_0,\eta_0))=0$, a conventional estimator of {$\theta_0$, say $\widehat{\theta}$}, would be the solution to 
$$
\sum_{j=1}^n u(Z_j;\theta,\widehat{\eta}) = 0,
$$
where $\widehat{\eta}$ is a consistent estimator of $\eta_0$ obtained in a preliminary step. However, it is well known that such a ``plug-in'' estimator is sensitive to the quality of the preliminary estimator $\widehat{\eta}$ used.

Assuming sufficient regularity, a standard argument based on a linearization around $\theta_0$ yields
$$
\left( \mathbb{E}\left(-\frac{\partial u(Z_j;\theta_0,\eta_0)}{\partial \theta}\right) + o_P(1) \right)
\,
(\widehat{\theta}-\theta_0)
=
\frac{1}{n}
\sum_{j=1}^n u(Z_j;\theta_0,\widehat{\eta}),
$$
so that the sampling properties of $\widehat{\theta}-\theta_0$ are dictated by the sampling properties of the estimating equation.
We have 
\begin{equation}\label{eq_expan_1}
\begin{split}
\frac{1}{n}
\sum_{j=1}^n u(Z_j;\theta_0,\widehat{\eta})
& =  
\underset{(A)}{\underbrace{\frac{1}{n} \sum_{j=1}^n u(Z_j;\theta_0,\eta_0)}}	
 \\
  &+
 \underset{(B)}{\underbrace{  \left(
 \frac{1}{n} \sum_{j=1}^n 
  \frac{\partial u(Z_j;\theta_0,\eta_0)}{\partial \eta}
  -
  \mathbb{E}\left( \frac{\partial u(Z_j;\theta_0,\eta_0)}{\partial \eta} \right)
  \right)
  \,
  \left(\widehat{\eta}-\eta_0 \right)}}
\\
& + 
  \underset{(C)}{\underbrace{\mathbb{E}\left( \frac{\partial u(Z_j;\theta_0,\eta_0)}{\partial \eta} \right)
  \,
  \left(\widehat{\eta}-\eta_0 \right)}}
\\
& 
+
 O_P(\lvert \widehat{\eta}-\eta_0 \rvert^{2}).
 \end{split}
\end{equation}

The (A) term in (\ref{eq_expan_1}) is a zero-mean sample average to which a standard central-limit theorem can be applied. Hence, it is generally $O_P(n^{-\nicefrac{1}{2}})$. The next two terms in the expansion capture the first-order effect of estimation noise in $\widehat{\eta}$. The (B) term can generally be ensured to be $o_P(n^{-\nicefrac{1}{2}})$. A generic approach to achieve this is to compute $\widehat{\eta}$ from data that are independent of $Z_1,\ldots, Z_n$, for example using sample splitting. In the case of (\ref{eq_expan_1}), (B) is the product of a sample average of zero-mean random variables---which is $O_P(n^{-\nicefrac{1}{2}})$---and an $o_P(1)$ term--- as $\widehat{\eta}$ is consistent for $\eta_0$---and, therefore, (B) is $o_P(n^{-\nicefrac{1}{2}})$.
The (C) term, however, features a non-random Jacobian that, in general, is non-zero. Hence, (C) is $O_P(\lvert\widehat{\eta}-\eta_0 \rvert)$, and will only be asymptotically negligible when $\widehat{\eta}$ is superconsistent for $\eta_0$, which is not usually the case.

Suppose now that $u$ is first-order orthogonal, in the sense that 
\begin{equation} \label{eq:neyman1}
  \mathbb{E}\left( \frac{\partial u(Z_j;\theta_0,\eta_0)}{\partial \eta} \right)
  =
  0.
\end{equation}
Then the (C) term vanishes from (\ref{eq_expan_1}) and we obtain
\begin{equation}
	\begin{split}
		\frac{1}{n}
		\sum_{j=1}^n u(Z_j;\theta_0,\widehat{\eta})
		= \frac{1}{n} \sum_{j=1}^n u(Z_j;\theta_0,\eta_0)	
		+
		O_P(\lvert\widehat{\eta}-\eta_0 \rvert^2)
		+ o_P(n^{-\nicefrac{1}{2}}).\label{eq_equ_1}
	\end{split}	
\end{equation}
The requirement that $\widehat{\eta}-\eta_0 = o_P(n^{-\nicefrac{1}{4}})$ then guarantees that the impact of the estimation error in $\widehat{\eta}$ on $\widehat{\theta}$ is asymptotically negligible. While a given function $u$ does not, in general, satisfy \eqref{eq:neyman1},  \cite{Neyman1959} proposed a general method to transform it into one that does. The resulting function is said to be  Neyman-orthogonal.

Condition \eqref{eq:neyman1} has a long history in semiparametric estimation problems (\citealp{Bickel1982}, \citealp{Schick1986}, \citealp{Newey1994}). More recently, it has proved to be a fundamental ingredient in the literature on high-dimensional inference (see \citealp*{ChernozhukovChetverikovDemirerDufloHansenNeweyRobins2018}).
There are, however, instances where it is ineffective. To illustrate this it suffices to consider the simple panel data setting from the \cite{NeymanScott1948} problem.

{\paragraph{Example: Neyman-Scott model (continued).} In this problem it is easy to verify that
$$
\mathbb{E}\left( \frac{\partial u(Y_i;\sigma_0^2,\eta_{i0})}{\partial \eta_i} \right)=-\frac{1}{\sigma_0^4}\sum_{j=1}^T\mathbb{E}\left(Y_{ij}-\eta_{i0}\right)=0,
$$ 
and so the score is already first-order Neyman-orthogonal with respect to the fixed effects. Nevertheless, given preliminary estimators $\widehat{\eta}_1,\ldots, \widehat{\eta}_N$, and letting $\nu_i = \widehat{\eta}_i - \eta_{i0}$, the estimator
\begin{equation}
\widehat{\sigma}^2 = 
\frac{1}{NT} \sum_{i=1}^N \sum_{j=1}^T (Y_{ij} - \widehat{\eta}_i)^2,\label{eq_sig2}
\end{equation}
has expectation
$
\sigma^2_0 -  \nicefrac{2}{N} \sum_{i=1}^N \mathbb{E}(\overline{\varepsilon}_i \, \nu_i)
+
\nicefrac{1}{N}\sum_{i=1}^N \mathbb{E}(\nu_i^2),
$
for $\overline{\varepsilon}_i = \nicefrac{1}{T} \sum_{j=1}^T \varepsilon_{ij}$.
Thus, when using sample splitting, the bias is $\nicefrac{1}{N}\sum_{i=1}^N \mathbb{E}(\nu_i^2)$, the mean squared error of the preliminary estimator. With cross-fitting this is, at best, $O(T^{-1})$. Hence,
$
\sqrt{NT} (\widehat{\sigma}^2 - \sigma^2_0)
$
will not have a correctly-centered limit distribution unless $\nicefrac{N}{T}\rightarrow 0$. However, under this condition, the joint maximum-likelihood estimator of $\sigma^2_0$ and the fixed effects, too, is asymptotically unbiased. Hence, having a score that is Neyman-orthogonal, even when combined with sample splitting, does not suffice to resolve the incidental parameter problem in panel data problems.
}

\subsection{Higher-order orthogonality\label{subsec_23}}

To see how Neyman-orthogonality to a higher order can be helpful we now consider a further expansion of (\ref{eq_expan_1}). Again assuming sufficient regularity, we have, for any integer $q\geq 1$, 
\begin{equation*} 
\begin{split}
\frac{1}{n}
\sum_{j=1}^n u(Z_j;\theta_0,\widehat{\eta})
& = \underset{(A)}{\underbrace{\frac{1}{n} \sum_{j=1}^n u(Z_j;\theta_0,\eta_0)}}
\\
& +
 \underset{(B)}{\underbrace{\sum_{p=1}^q
 \frac{1}{p!}
 \left(
 \frac{1}{n} \sum_{j=1}^n 
  \frac{\partial^p u(Z_j;\theta_0,\eta_0)}{\partial \eta^p}
  -
  \mathbb{E}\left( \frac{\partial^p u(Z_j;\theta_0,\eta_0)}{\partial \eta^p} \right)
  \right)
  \left(\widehat{\eta}-\eta_0 \right)^p}}
\\
& + 
\underset{(C)}{\underbrace{\sum_{p=1}^q 
 \frac{1}{p!}
  \mathbb{E}\left( \frac{\partial^p u(Z_j;\theta_0,\eta_0)}{\partial \eta^p} \right)
  \,
  \left(\widehat{\eta}-\eta_0 \right)^p}}
\\
& 
+
 O_P(\lvert \widehat{\eta}-\eta_0 \rvert^{q+1}).
\end{split}
\end{equation*}
Here, the (A) term is the same as before. Also, with sample splitting we can again ensure that the (B) term will be asymptotically negligible. On the other hand, if the function $u$ satisfies the higher-order orthogonality condition
\begin{equation} \label{eq:neymanq}
  \mathbb{E}\left( \frac{\partial^p u(Z_j;\theta_0,\eta_0)}{\partial \eta^p} \right)
  =
  0
  ,\qquad
  1\leq p \leq q,
\end{equation}
the (C) term is equal to zero, and so
\begin{equation}
\begin{split}
\frac{1}{n}
\sum_{j=1}^n u(Z_j;\theta_0,\widehat{\eta})
 = \frac{1}{n} \sum_{j=1}^n u(Z_j;\theta_0,\eta_0)	
 +
O_P(\lvert\widehat{\eta}-\eta_0 \rvert^{q+1})
 + o_P(n^{-\nicefrac{1}{2}}).
\end{split}	\label{eq_equ_2}
\end{equation}	
Comparing (\ref{eq_equ_2}) to (\ref{eq_equ_1}) we see that the impact of estimation noise in $\widehat{\eta}$ on our estimator of $\theta_0$ has been reduced further. Moreover, for the impact of estimation error to be negligible, we now only require that $\lvert \widehat{\eta}-\eta_0 \rvert^{q+1} = o_P(n^{-\nicefrac{1}{2}})$. It then follows from standard results that, as $n\rightarrow\infty$,
$$
\sqrt{n}(\widehat{\theta}-\theta_0) \overset{d}{\rightarrow}  {\cal N}(0,\Sigma_\theta)
$$
for some $\Sigma_\theta$, provided that
$$
\widehat{\eta}-\eta_0 = o_P(n^{-\nicefrac{1}{2(q+1)}}).
$$

The notion of $q$th-order Neyman-orthogonality as in (\ref{eq:neymanq}) was introduced by \citet{MackeySyrgkanisZadik2018}. In the context of our conditional-likelihood setup, we will give a general procedure to construct higher-order Neyman-orthogonal functions below.

\paragraph{Example: Neyman-Scott model (continued).} In the model of \cite{NeymanScott1948},
	$$\mathbb{E}\left( \frac{\partial^2 u(Y_i;\sigma_0^2,\eta_{i0})}{\partial \eta_i^2} \right)=\frac{T}{\sigma_0^4}\neq 0.$$ It thus follows that $u$ is not orthogonal to second order. Below we will show that a second-order Neyman-orthogonal score equation exists. This estimating equation does not involve the nuisance parameters at all -- in particular, it does not depend on the choice of preliminary estimator of $\eta_{i0}$ -- and its solution is the usual degrees-of-freedom corrected estimator
 \begin{equation}
 \widehat{\sigma}^2
 =
 \frac{1}{N(T-1)} \sum_{i=1}^N \sum_{j=1}^T (Y_{ij}-\overline{Y}_i)^2,\label{eq_sig2_hat}
\end{equation}
 where $\overline{Y}_i = \nicefrac{1}{T} \sum_{j=1}^T Y_{ij}$. The estimator $\widehat{\sigma}^2$ is well-known to be fixed-$T$ consistent.

\renewcommand{\theequation}{\arabic{section}.\arabic{equation}}  \setcounter{equation}{0}

\section{Estimation based on orthogonalized functions\label{sec_est}}

We now present our estimation approach in the general case where the target parameter $\mu_0$ may be equal to $\theta_0$ or may be a different parameter such as an average effect, and there are multiple, vector-valued nuisance parameters $\eta_{i0}$. We start by formally defining higher-order Neyman-orthogonality in this general setup  and describe estimation based on higher-order Neyman-orthogonal moment functions. In the next section, we will then show how to construct such functions.

\subsection{Definition of higher-order orthogonality}
\label{subsec:DefHigherOrder}

Let $d_\eta$ be the dimension of $\eta$ and write $\eta=(\eta_1,\ldots,\eta_{d_\eta})$. For any non-negative integer $p$ and a vector of integers $m = (m_1,\ldots, m_p)$ satisfying $1\leq m_s \leq d_\eta$ for all $1\leq s\leq p$, define 
\begin{equation}
D^m_{\eta}  = 
\frac{\partial^p }{\partial \eta_{m_1}\cdots \partial \eta_{m_p}}. \label{DefGenPartialDerivative}
\end{equation}
For a given $p$, there are $
d_p = \binom{d_\eta+p-1}{p}
$
unique such partial derivatives. Let
$
\nabla^{(p)}_\eta
$
be the vector operator of dimension $d_p$ that collects all these unique partial derivatives of order $p$. Finally, let $\nabla^q_\eta$ be the vector operator of dimension $\sum_{p=1}^q d_p$ obtained on stacking $\nabla^{(p)}_\eta$ for $p=1,\ldots,q$. Explicitly, we have
$$
 \nabla^{q}_\eta
   = \left(\begin{array}{c}
           \nabla^{(1)}_\eta \\
            \nabla^{(2)}_\eta \\
           \vdots \\
           \nabla^{(q)}_\eta
      \end{array} \right)
   =    \left(\begin{array}{c}
           \left[D^m_{\eta}  \, :\, m \in \{1,\ldots,d_\eta\} \right] 
           \\
            \left[D^m_{\eta}  \, :\, m \in \{1,\ldots,d_\eta\}^2, \,
            m_1 \leq m_2\right]
            \\
           \vdots \\
           \left[D^m_{\eta} \, :\,  m \in \{1,\ldots,d_\eta\}^q, \, m_1 \leq m_2 \leq \ldots \leq m_q\right]
      \end{array} \right) .
$$

Neyman-orthogonality to order $q$ can now be defined as follows
(\citealp{MackeySyrgkanisZadik2018}).

\begin{definition}\label{def_ortho}
If the function $u$ satisfies 
\begin{align}
\mathbb{E}\left[ \nabla^q_\eta \, u(Z;\theta_0,\eta_0,\mu_0) \right] = 0,
     \label{NeymanOrthDef}
\end{align}
for some integer $q$, then we say that $u$ is Neyman-orthogonal to order $q$.
\end{definition}

\noindent
In this definition, {\it all} possible partial derivatives of $u(Z;\theta,\eta,\mu)$ with respect to $\eta$ up to order 
$q$ have mean zero.

\subsection{Estimation}
\label{subsect:Estimation}

Let $\mu_0$ satisfy (\ref{MainMoment}) for a (possibly vector-valued) function $u$. We assume that $u$ is Neyman-orthogonal to order $q$ with respect to $\eta_i$, in the sense of Definition \ref{def_ortho}. Suppose that we have access to preliminary estimators $\widehat{\eta}_1,\ldots, \widehat{\eta}_N$ of the nuisance parameters that are independent of the data $Z_1,\ldots,Z_N$. If $\eta_{i0}$ is defined as the solution to a moment condition involving the same data, estimation based on sample-splitting, combined with cross-fitting (see, e.g., \citealp{NeweyRobins2017}), can be applied. When the observations are independent this is conventional. For situations where the data are  dependent, modified sample-splitting strategies are available (see, e.g., \citealp{semenova2023inference}).

We estimate $\mu_0$ by the GMM estimator
\begin{equation}
\widehat{\mu}
=
\underset{\mu}{\mbox{argmin}}\, \left\|\sum_{i=1}^Nu(Z_i;\widehat{\theta},\widehat{\eta}_i,\mu)\right\|_{W},\label{muEstimation}
\end{equation}
where $W$ is a chosen symmetric positive-definite matrix, $\lVert u \rVert_W = \sqrt{u^\top W \, u}$, and $\widehat{\theta}$ is an estimator of $\theta_0$.

The estimator $\widehat{\theta}$ will depend on the problem at hand. If $\theta_0$ is defined through a moment condition of the form $\sum_{i=1}^N\mathbb{E}(\widetilde{u}(Z_i;\theta_0,\eta_{i0})) = 0$, for a function $\widetilde{u}$ that is Neyman-orthogonal to order $q$, then our framework can be applied and we can use
\begin{equation}\widehat{\theta}=\underset{\theta}{\mbox{argmin}}\, \left\|\sum_{i=1}^N\widetilde{u}(Z_i;\theta,\widehat{\eta}_i)\right\|_{\widetilde{W}},\label{thetaEstimation}\end{equation}
where $\widetilde W$ is again a chosen weight matrix. In this case, we may equally combine \eqref{muEstimation} and \eqref{thetaEstimation} into a single GMM estimation procedure.

In Section \ref{sec_asympt}, we provide conditions under which this approach yields estimators that are $n^{-\nicefrac{1}{2}}$-consistent and asymptotically normal, where $n$ is the total number of observations. We will impose two key conditions. The first one is that, although their number may increase with the sample size, the dimension of each $\eta_i$ remains bounded as $n$ tends to infinity. This imposes a suitable sense of sparsity in the relationship between the nuisance parameters and the outcomes. This condition is trivially satisfied in the panel data and network problems with fixed effects that we consider. The second key condition we impose is that the convergence rates of the preliminary estimates $\widehat{\eta}_i$ be faster than $n^{-\nicefrac{1}{2(q+1)}}$. This ensures that, after having orthogonalized to order $q$, any remainder terms are asymptotically negligible.

Our approach requires choosing an orthogonality order $q$. In practice, we recommend reporting estimates and standard errors for various values $q=1,2,...,q_{\rm max}$, each based on a $q$-orthogonal function $u_q^*$. For a given $q$ that satisfies the conditions of our theory (see Theorem \ref{th:Asymptotic} in Section \ref{sec_asympt}), the estimators based on $u_q^*$ and $u_{q+1}^*$ are both root-$n$ consistent and asymptotically normal. Given this, we propose to report as a diagnostic the difference in orthogonal moment functions $u^*_{q}-u^*_{q+1}$, suitably normalized. A large value of this statistic indicates that the order $q$ of orthogonality may be too small. We describe this approach in Appendix \ref{App_q}, while leaving the formal construction of a selection method for $q$ and its impact on inference on $\mu_0$ to future work.

\renewcommand{\theequation}{\arabic{section}.\arabic{equation}}  \setcounter{equation}{0}

\section{Achieving higher-order Neyman-orthogonality\label{sec_Construct}}

{
\subsection{Main result}

Let $u$ be a moment function. We now show how to construct an orthogonalized counterpart of $u$, which we call $u_q^*$, that is Neyman-orthogonal to order $q$, where $q\geq 1$ is any arbitrary order. 

Recall the definition of the vector operators $\nabla^{(p)}_\eta$
and $\nabla^{q}_\eta$ in Subsection~\ref{subsec:DefHigherOrder}. It is convenient to
introduce the \cite{Bhattacharyya1946} basis $v_1,v_2,\ldots$, where
$$
v_p(z;\theta,\eta) = \frac{\nabla_{\eta}^{(p)} \ell(y\,|\,  x;\theta,\eta)}{\ell(y\,|\,  x;\theta,\eta)}.
$$
 \cite{SmallMcLeish1994} discuss several properties of this basis. One important property for our purposes is that
\begin{equation} \label{eq:zeromean}
\mathbb{E}_{\theta,\eta}(v_p(Z;\theta,\eta) \,|\,  X=x)
=
\int v_p(z;\theta,\eta) \, \ell(y\,|\,  x;\theta,\eta) \, dy = 0
\end{equation}
for any $p$, so all elements of the Bhattacharyya basis have (conditional) mean equal to zero. In (\ref{eq:zeromean}), and throughout this section, $\mathbb{E}_{\theta,\eta}(\cdot \,|\,  X=x) $ denotes the conditional expectation under $\ell(y\,|\,  x;\theta,\eta)$.

The low-order basis functions are familiar from likelihood theory. For example,
\begin{equation*}
\begin{split}
v_1(z;\theta,\eta) 
& =
\ \
 \frac{\partial \log \ell(y\,|\,  x;\theta,\eta)}{\partial \eta},
 \\
 v_2(z;\theta,\eta) 
&  =
\frac{\partial \log \ell(y\,|\,  x;\theta,\eta)}{\partial \eta}\frac{\partial \log \ell(y\,|\,  x;\theta,\eta)}{\partial \eta^{\top}}
+
\frac{\partial^2 \log \ell(y\,|\,  x;\theta,\eta)}{\partial \eta\partial \eta^{\top}}.
 \end{split}
\end{equation*}
The fact that these functions have mean zero follows from the unbiasedness of the score and from the information equality, respectively.

Stacking the leading $q$ basis functions, we obtain
$$
w_q(z;\theta,\eta) =\frac{\nabla_{\eta}^q \ell(y\,|\,  x;\theta,\eta)}{ \ell(y\,|\,  x;\theta,\eta)}
 = 
 \left( \begin{array}{c}
       v_1(z;\theta,\eta) \\
       \vdots  \\
       v_q(z;\theta,\eta)
  \end{array}
 \right). 
$$ 
The vectors $w_q$ are mean-zero ``generalized score functions''. The vector space spanned by the Bhattacharyya basis at order $q$ is a higher-order tangent set.\footnote{In models with i.i.d. observations, for which the likelihood of any $q$ observations is $\prod_{r=1}^q\ell(z_r)$, the tangent set of order $q$ is defined as the span of all derivatives with respect to the parameter, up to order $q$, of $\prod_{r=1}^q\ell(z_r)$, divided by $\prod_{r=1}^q\ell(z_r)$; see equation (1.8) in \citet{vanderVaart2014}. In our setting we do not assume that observations are i.i.d.} While it is possible to achieve higher-order Neyman-orthogonality using other bases of functions, the Bhattacharyya basis delivers simple expressions through the use of Bartlett identities.

Next, let us define the matrices 
$$
\varSigma_{w_qw_q}(x;\theta,\eta)
=
\mathbb{E}_{\theta,\eta}(w_q(Z;\theta,\eta)\, w_q(Z;\theta,\eta)^\top\,|\,  X=x) ,
$$
and
$$
\varSigma_{w_qu}(x;\theta,\eta,\mu)
=
\mathbb{E}_{\theta,\eta}(w_q(Z;\theta,\eta)\, u(Z;\theta,\eta,\mu)^\top\,|\,  X=x),
$$
which are, respectively, the (conditional) covariance matrix of the first $q$ members of the Bhattacharyya basis, and the covariance matrix
of the same $q$ basis functions with the vector function $u$.
Finally, let
$$
b_q(x;\theta,\eta,\mu) = 
\nabla_{\eta}^q \,
\mathbb{E}_{\theta,\eta}(u(Z;\theta,\eta,\mu)^\top \vert X=x)
.
$$
Note that $b_q$ is zero when $u$ is the score for $\theta$, i.e., 
$\frac{\partial \log \ell(y\,|\,  x;\theta,\eta)}{\partial\theta}$. In general, however, $b_q$ will be non-zero.
Here we
assume that $u(z;\theta,\eta,\mu)$ and $\ell(y\,|\,  x;\theta,\eta)$
     are sufficiently often differentiable in~$\eta$, and that the expectations in the definitions of
     $\varSigma_{w_qw_q}$,
     $\varSigma_{w_qu}$,
     and $b_q$ are well-defined.

The proof of the following result is in Appendix \ref{App_proofs}.

\begin{theorem}\label{theo_neyman}
Suppose that $\varSigma_{w_qw_q}(x;\theta,\eta)$ is invertible and
let
$$
A(x;\theta,\eta,\mu)
=
\varSigma_{w_qw_q}(x;\theta,\eta)^{-1} 
\left(
\varSigma_{w_q u}(x;\theta,\eta,\mu)
-
b_q(x;\theta,\eta,\mu)
\right).
$$
Then the function
$$
u_q^*(z;\theta,\eta,\mu)
=
u(z;\theta,\eta,\mu)
-
A(x;\theta,\eta,\mu)^\top \, 
w_q(z;\theta,\eta)
$$
satisfies
$\mathbb{E}_{\theta,\eta} (  \nabla^q_\eta \, u_q^*(Z;\theta,\eta,\mu)  \, \vert \, X=x ) = 0$.
This implies that $u_q^*$
is Neyman-orthogonal to order $q$, as defined above.

\end{theorem}

}

\noindent
Theorem \ref{theo_neyman} generalizes the projected-score construction of \citet{SmallMcLeish1989} and \citet{WatermanLindsay1996}. To see this, consider the case where $\mu_0=\theta_0$, and $u$ is the score function for $\theta$. Then $b_q=0$ and Theorem \ref{theo_neyman} yields
$$
u_q^*(z;\theta,\eta,\mu)
=
u(z;\theta,\eta,\mu)
-\left(\varSigma_{w_qw_q}(x;\theta,\eta)^{-1} 
\varSigma_{w_q u}(x;\theta,\eta)\right)^\top \, 
w_q(z;\theta,\eta),
$$
which is the projected score of order $q$. The projected score was originally developed as a tool to achieve E-ancillarity (\citealp{SmallMcLeish1988}) and to approximate the conditional score for $\theta$, when the latter exists (\citealp{WatermanLindsay1996}). It generalizes \cite{Neyman1959} in that $u_q^*$ is the (population) residual of a least-squares regression of $u$ on $w_q$; thus,
$$
\mathbb{E}_{\theta,\eta}
(
w_q(Z;\theta,\eta)
\,
u_q^*(Z;\theta,\eta,\mu)^\top
\vert X=x)
=0.
$$
While the fact that $u_q^*$ is Neyman-orthogonal is noted by \citet{WatermanLindsay1996} (although a link with Neyman's work is not made), it is not exploited. Moreover, unlike the conditional score, the projected score still depends on $\eta$, and it will generally not have improved properties over the score itself. As we highlight here, it is the combination of higher-order versions of Neyman-orthogonality with sample splitting that allows one to improve over working with the original score.

Theorem \ref{theo_neyman} covers more general estimating equations as well as more general parameters of interest, such as average elasticities or counterfactual quantities. Incorporating $b_q(x;\theta,\eta,\mu)$ into $u_q^*(z;\theta,\eta,\mu)$ is a key innovation that enables this generality. Observe that, in this case, we have that
$$
\mathbb{E}_{\theta,\eta}
(
w_q(Z;\theta,\eta)
\,
u_q^*(Z;\theta,\eta,\mu)^\top
\vert X=x)
=
b_q(x;\theta,\eta,\mu),
$$
thereby revealing $u_q^*$ to be an influence function of order $q$ per Equation (1.9) in \cite{vanderVaart2014}.

We note that Theorem \ref{theo_neyman} requires the matrix $\varSigma_{w_qw_q}(x;\theta,\eta)$ to be invertible. In the standard case of first-order Neyman-orthogonality this corresponds to non-singularity of the information matrix of the nuisance parameters. For higher-order Neyman-orthogonality this requirement imposes further restrictions. For example, in the team production model (\ref{eq_CES_size2}) with two-worker teams, the $2\times 2$ matrix $
\varSigma_{w_1w_1}({\cal{K}};\theta,\eta)$ is not invertible, as only the sum $\eta_{k(j,1)}^{\gamma}+\eta_{k(j,2)}^{\gamma}$ can be identified. In our application in Section \ref{sec_appli}, we will tackle this issue by combining data on teams of size 2 with single-author articles, and working with subsets $i$ of three teams each.\footnote{As another example, consider the standard binary-choice panel data model 
$$
\mathbb{P}_{\theta,\eta_i}(Y_{ij} = 1 \,|\,  X_{i1},...,X_{iT})
=
\Phi(\eta_i + X_{ij}^\top \theta)
,\quad i=1,\ldots,N,\quad j=1,\ldots,T,
$$
for (conditionally-independent) binary outcomes $Y_{ij}$ and covariates $X_{ij}$. In this model, the rank of $\varSigma_{w_qw_q}(x;\theta,\eta)$ is bounded by $2^T-1$, so $\varSigma_{w_qw_q}(x;\theta,\eta)$ is singular for all $q\geq 2^T$.}

\subsection{Intuition and discussion\label{subsec_intuition}}

\noindent
To gain intuition into the construction in Theorem \ref{theo_neyman} it is useful to again consider the case where $u$ is a univariate function, the nuisance parameter is a scalar, and one wishes to estimate $\mu_0=\theta_0$. Throughout this subsection we also assume that the conditional mean $\mathbb{E}_{\theta,\eta}\left(u(Z;\theta,\eta)\,|\, X=x\right)$ does not vary with $\eta$, so that the term $b_q(x;\theta,\eta,\mu)$ in Theorem \ref{theo_neyman} is zero; this holds, for example, when $u$ is the score for $\theta$. The general case is analogous, with the $b_q$ term retained.

\paragraph{First-order orthogonality.}
To relate our approach to the literature consider first $q=1$. Let
\begin{equation}
u_1^*(z;\theta,\eta) = u(z;\theta,\eta) - a_1(x;\theta,\eta) \, v_1(z;\theta,\eta),\label{eq_ustar}
\end{equation}
for some function $a_1$. Note that, by virtue of \eqref{eq:zeromean}, the term involving $v_1$ does not introduce any bias. We have
$$
\frac{\partial u_1^*(z;\theta,\eta)}{\partial\eta} =
\frac{\partial u(z;\theta,\eta)}{\partial\eta}
-
\frac{\partial  a_1(x;\theta,\eta)}{\partial\eta}
\,
v_1(z;\theta,\eta)
-
a_1(x;\theta,\eta)
\,
\frac{\partial v_1(z;\theta,\eta)}{\partial\eta}.
$$
Take conditional expectations and exploit \eqref{eq:zeromean} to see that
$$
\mathbb{E}_{\theta,\eta}
\left(
\left.
\frac{\partial u_1^*(Z;\theta,\eta)}{\partial\eta}
\right\rvert X=x
\right) 
=
0
$$
if and only if 
$$
\mathbb{E}_{\theta,\eta}
\left(
\left.
\frac{\partial u(Z;\theta,\eta)}{\partial\eta}
\right\rvert X=x
\right)
-
a_1(x;\theta,\eta) \
\mathbb{E}_{\theta,\eta}
\left(
\left. 
\frac{\partial v_1(Z;\theta,\eta)}{\partial\eta}
\right\rvert X=x
\right) = 0.
$$
This is achieved by setting
\begin{equation}
a_1(x;\theta,\eta)
=
\left(
\mathbb{E}_{\theta,\eta}
\left(
\left. 
\frac{\partial v_1(Z;\theta,\eta)}{\partial\eta}
\right\rvert X=x
\right)
\right)^{-1}
\mathbb{E}_{\theta,\eta}
\left(
\left.
\frac{\partial u(Z;\theta,\eta)}{\partial\eta}
\right\rvert X=x
\right).\label{eq_a1}
\end{equation}
Iterating expectations  shows that the resulting function $u_1^*$ is Neyman-orthogonal to order $q=1$. By the information matrix equality we have
$$\mathbb{E}_{\theta,\eta}
\left(
\left. 
\frac{\partial v_1(Z;\theta,\eta)}{\partial\eta}
\right\rvert X=x
\right)=-\mathbb{E}_{\theta,\eta}
\left(
\left. 
 v_1(Z;\theta,\eta)^2
\right\rvert X=x
\right)=-\varSigma_{w_1w_1}(x;\theta,\eta),$$
and 
$$\mathbb{E}_{\theta,\eta}
\left(
\left.
\frac{\partial u(Z;\theta,\eta)}{\partial\eta}
\right\rvert X=x
\right)=-\mathbb{E}_{\theta,\eta}
\left(
\left. 
 v_1(Z;\theta,\eta)u(Z;\theta,\eta)
\right\rvert X=x
\right)=-\varSigma_{w_1u}(x;\theta,\eta),
$$
leading to the representation of the function $u_1^*$ as in the theorem.

The above derivation of \eqref{eq_a1} is well-known. Furthermore, it does not hinge on the likelihood structure. Indeed, recent work exploiting orthogonality, such as that surveyed in \cite*{ChernozhukovChetverikovDemirerDufloHansenNeweyRobins2018},  does so in the context of moment conditions. In our setup, as in \citeauthor{Neyman1959}'s (\citeyear{Neyman1959}) original work, the likelihood structure implies that $a_1$ is known up to the model parameters $\theta$ and $\eta$ (conditional on the regressors). Outside of this framework, in contrast, $a_1$ needs to be treated as an additional nuisance parameter. This is possible because, as $u_1^*$ is linear in $a_1$, it is automatically first-order Neyman-orthogonal to it by virtue of \eqref{eq:zeromean}. This logic, however, does not extend to higher order, as the implied system of equations becomes inconsistent, so that no solution exists, as we will see next.

\paragraph{Higher-order orthogonality.}
Let $q=2$, and again
consider a linear transformation of $u$, now involving the leading two Bhattacharyya basis functions. This gives
\begin{equation}u_{2}^*(z;\theta,\eta)
=
u(z;\theta,\eta)
-
\left( 
\begin{array}{c}
a_{21}(x;\theta,\eta)
\\
a_{22}(x;\theta,\eta)
\end{array}
\right)^\top
\,
\left( 
\begin{array}{c}
v_1(z;\theta,\eta)
\\
v_2(z;\theta,\eta)
\end{array}
\right).\label{eq_u2_star}
\end{equation}
Taking first-derivatives with respect to the nuisance parameter, and proceeding as in the first-order case, gives
$$
\mathbb{E}_{\theta,\eta}
\left(
\left.
\frac{\partial u(Z;\theta,\eta)}{\partial\eta}
\right\rvert X=x
\right)
=
\left( 
\begin{array}{c}
 a_{21}(x;\theta,\eta)
\\
a_{22}(x;\theta,\eta)
\end{array}
\right)^\top
\,
\left( 
\begin{array}{c}
\mathbb{E}_{\theta,\eta}
\left(
\left.
\frac{\partial v_1(Z;\theta,\eta)}{\partial\eta}
\right\rvert X=x
\right)
\\
\mathbb{E}_{\theta,\eta}
\left(
\left.
\frac{\partial v_2(Z;\theta,\eta)}{\partial\eta}
\right\rvert X=x
\right)
\end{array}
\right).
$$
Solving this equation for $a_{21}$ for given $a_{22}$ yields
\begin{equation}\label{eq_a21}
a_{21}(x;\theta,\eta)
=
a_1(x;\theta,\eta) -  c_1(x;\theta,\eta) \, a_{22}(x;\theta,\eta),
	\end{equation}
where $a_1$ is given by (\ref{eq_a1}) and 
$$
c_1(x;\theta,\eta)
=
\left(
\mathbb{E}_{\theta,\eta}
\left(
\left.
\frac{\partial v_1(Z;\theta,\eta)}{\partial\eta}
\right\rvert X=x
\right)
\right)^{-1}
\mathbb{E}_{\theta,\eta}
\left(
\left.
\frac{\partial v_2(Z;\theta,\eta)}{\partial\eta}
\right\rvert X=x
\right).
$$
The coefficient $c_1$ has the same form as $a_1$, except that it features $v_2$ instead of $u$. Moreover, 
plugging (\ref{eq_a21}) back into (\ref{eq_u2_star}) yields
\begin{equation*}
u_{2}^*(z;\theta,\eta)
=
u_{1}^*(z;\theta,\eta) - a_{22}(x;\theta,\eta) \, v_2^*(z;\theta,\eta),
\end{equation*}
where
$
v_2^*(z;\theta,\eta)
=
v_2(z;\theta,\eta)
-
c_1(x;\theta,\eta) \,
v_1(z;\theta,\eta).
$
Note that $v_2^*$ is Neyman-orthogonal to order 1, that is,
$$
\mathbb{E}_{\theta,\eta}\left(\left. \frac{\partial v_2^*(Z;\theta,\eta)}{\partial\eta}\right\rvert X=x\right) = 0.
$$ It follows that $u_2^*$ is Neyman-orthogonal to order 1 for any $a_{22}$. We will now choose $a_{22} $ such that $u_2^*$ is Neyman-orthogonal to order 2.

Next, differentiating $u_2^*$ with respect to $\eta$ twice gives
\begin{equation*}
\begin{split}
\frac{\partial^2 u_2^*(z;\theta,\eta)}{\partial\eta^2}
& =
\frac{\partial^2 u_1^*(z;\theta,\eta)}{\partial\eta^2}
-
a_{22}(x;\theta,\eta) \,
\frac{\partial^2 v_2^*(z;\theta,\eta)}{\partial\eta^2}
\\
& -
\frac{\partial^2 a_{22}(x;\theta,\eta)}{\partial\eta^2} 
\,
v_2^*(z;\theta,\eta)
-
2 
\frac{\partial a_{22}(x;\theta,\eta)}{\partial\eta} 
\,
\frac{\partial v_2^*(z;\theta,\eta)}{\partial\eta}.
\end{split}
\end{equation*}
Since $v_2^*$ has zero mean and is orthogonal to order 1, the terms involving the first and second derivative of $a_{22}$ drop out when taking expectations. It follows that $u_2^*$ in (\ref{eq_u2_star}) is Neyman-orthogonal to order 2 when one sets $a_{21}$ to its expression in (\ref{eq_a21}), and $a_{22}$ to
\begin{equation}
a_{22}(x;\theta,\eta)
=
\left(
\mathbb{E}_{\theta,\eta}
\left(
\left. 
\frac{\partial^2 v_2^*(Z;\theta,\eta)}{\partial\eta^2}
\right\rvert X=x
\right)
\right)^{-1}
\mathbb{E}_{\theta,\eta}
\left(
\left.
\frac{\partial^2 u_1^*(Z;\theta,\eta)}{\partial\eta^2}
\right\rvert X=x
\right).\label{eq_a22}
\end{equation}
Note that this construction amounts to solving a system of linear equations. The fact that the solution in (\ref{eq_a21})--(\ref{eq_a22}) coincides with the expression in Theorem \ref{theo_neyman} may then again be verified by using Bartlett identities.

To appreciate the role of the likelihood structure in the above argument, suppose that $a_{21}$ and $a_{22}$ are not known up to the parameters $\theta$ and $\eta$. Then they are additional nuisance parameters and, thus, we require that all first- and second-order derivatives with respect to $(a_{21},a_{22})$, and $\eta$ be mean zero. The cross-derivatives between $(a_{21},a_{22})$ and $\eta$ are problematic, since having those to be mean zero would require
$$
\mathbb{E}_{\theta,\eta}\left(\left. \frac{\partial v_1(Z;\theta,\eta)}{\partial \eta} \right\vert X=x \right)=0,
\qquad
\mathbb{E}_{\theta,\eta}\left(\left. \frac{\partial v_2(Z;\theta,\eta)}{\partial \eta} \right\vert X=x \right) =0,
$$
which is not generally the case.

\renewcommand{\theequation}{\arabic{section}.\arabic{equation}}  \setcounter{equation}{0}

\section{Examples\label{sec_ex}}

\subsection{Panel data models\label{subsec_panel}}
Consider an $N\times T$ panel data model with individual effects. Here, the likelihood factors across the cross-sectional observations and the likelihood contribution of unit $i$ takes the form
$$
\prod_{j=1}^T   f(Y_{ij} \,|\, X_{ij}; \theta, \eta_{i}).
$$
The maximum-likelihood estimator is well-known to suffer from a bias that is $O(T^{-1})$; see \cite{HahnNewey2004} and \cite{HahnKuersteiner2011} for derivations of this bias in static and dynamic models, respectively. Consider the estimation of $\theta_0$. The bias in the estimator comes from bias in the score stemming from estimation noise in the fixed effects. Taking $\eta_{i}$ to be scalar for notational simplicity, and letting $\widehat{\eta}_i$ be an estimator of $\eta_{i0}$, an expansion of the (normalized) score\footnote{In this discussion we work with the score divided by $T$, to facilitate the comparison with the panel data literature.} 
$$
u(Z_i;\theta_0,\widehat{\eta}_i)
{=}
\frac{1}{T} \sum_{j=1}^T
\frac{\partial \log f(Y_{ij} \vert X_{ij};\theta_0,\widehat{\eta}_i)}{\partial\theta}
$$
yields
\begin{align*}
u(Z_i;\theta_0,\widehat{\eta}_i)=&
u(Z_i;\theta_0,\eta_{i0})
+
\frac{\partial u(Z_i;\theta_0,\eta_{i0})}{\partial \eta_{i}}
\,
(\widehat{\eta}_i - \eta_{i0})
+\frac{1}{2}
\frac{\partial^2 u(Z_i;\theta_0,\eta_{i0})}{\partial \eta_i^2}
\,
(\widehat{\eta}_i - \eta_{i0})^2\\&+o_P(\lvert\widehat{\eta}_i - \eta_{i0} \rvert^2).
\end{align*}
Taking expectations and re-arranging shows that
\begin{equation*}
\begin{split}
\mathbb{E}
(u(Z_i;\theta_0,\widehat{\eta}_i))
 =&
\mathrm{cov}
\left(
\frac{\partial u(Z_i;\theta_0,\eta_{i0})}{\partial \eta_{i}},
\widehat{\eta}_i - \eta_{i0}
\right)
\\
& +
\mathbb{E}
\left(
\frac{\partial u(Z_i;\theta_0,\eta_{i0})}{\partial \eta_{i}}
\right)
\,
\mathbb{E}
(\widehat{\eta}_i - \eta_{i0})
\\
& +
\frac{1}{2}
\mathbb{E}
\left(
\frac{\partial^2 u(Z_i;\theta_0,\eta_{i0})}{\partial \eta_i^2}
\right)
\,
\mathbb{E} ((\widehat{\eta}_i - \eta_{i0})^2)
+
o(\mathbb{E}(\lvert \widehat{\eta}_i - \eta_{i0} \rvert^2)).
\end{split}
\end{equation*}
If we set $\widehat{\eta}_i = \widehat{\eta}_i(\theta_0) = \arg\max_{\eta} \sum_{j=1}^T \log f(Y_{ij} \,|\,  X_{ij}; \theta_0, \eta)$, the maximum-likelihood estimator (MLE) given $\theta_0$, then each one of these terms is $O(T^{-1})$. If we use an estimator $\widehat{\eta}_i$ that is independent of the estimation sample, the first term disappears. However, the remaining terms, which capture the nonlinearity bias and variance in the estimator of $\eta_{i0}$, remain. 
\cite{HahnNewey2004}, \cite{ArellanoHahn2007}, and \cite{DhaeneJochmans2015b,DhaeneJochmans2015a} present estimators of these terms based on the MLE that can be used to construct a bias-corrected estimator.

\cite{Lancaster2002} and \cite{woutersen2002robustness} integrate-out the fixed effects using a uniform prior after orthogonalizing to order 1 to obtain an estimator with bias $o(T^{-1})$;   \cite{Arellano2003} presents an alternative derivation of the same result. First-order Neyman-orthogonality, by itself, does not suffice as it does not handle the third term in the expansion,
that is, it does not properly correct for the noise in the estimated fixed effects. \cite{LiLindsayWaterman2003}, building on \cite{WatermanLindsay1996}, show that their (second-order) projected score for $\theta$, when evaluated at $\widehat{\eta}_i(\theta)$, is a first-order unbiased estimating equation for $\theta$. Thus, here, a sample-splitting procedure is not needed to achieve bias reduction. This is a consequence of the (second- or higher-order) projected score being orthogonal to the influence function of $\widehat{\eta}_i(\theta)$. While interesting, it is not clear whether this property extends to higher-order projections or to other parameters of interest, such as average marginal effects.

More generally, with $\widehat{\eta}_i - \eta_{i0} = O_P(T^{-\nicefrac{1}{2}})$, the score admits a higher-order expansion of the form,
$$
\mathbb{E}
(u(Z_i;\theta_0,\widehat{\eta}_i))
=
\frac{B_1}{T}
+
\frac{B_2}{T^2} + \cdots + \frac{B_q}{T^q} + o(T^{-q})
$$
for constants $B_1,B_2,\ldots, B_q$.
The maximum-likelihood estimator has $B_1\neq 0$, in general, and so requires that $\nicefrac{N}{T}\rightarrow 0$ to be asymptotically unbiased. The approaches to bias correction mentioned above remove $B_1$ but not the remaining terms. Approaches that estimate and subsequently remove all $B_p$, $1\leq p \leq q$, to enable asymptotically valid inference are given by \cite{DhaeneJochmans2015b,DhaeneJochmans2015a}. Relatedly, an estimator based on Neyman-orthogonalization, combined with a sample-splitting estimator that uses preliminary estimators that satisfy $\widehat{\eta}_i - \eta_{i0} = O_P(T^{-\nicefrac{1}{2}})$, renders the asymptotic bias negligible provided $N=o(T^q)$. The two approaches are not equivalent, however. Removing $B_1,\ldots,B_q$ leaves a bias of order $o(T^{-q})$, which permits asymptotically-valid inference under the weaker requirement that $N=o(T^{2q-1})$. Our rate condition $N=o(T^{q})$, which follows from Assumption \ref{ass:asymptotic1}$(v)$ below, is more restrictive as soon as $q\geq 2$. In exchange, our approach does not require characterizing, estimating, or removing the individual bias terms $B_p$.

\paragraph{Example: Neyman-Scott model (continued).}
Recall that the (un-normalized) unit-specific score for $\sigma^2$ is given by (\ref{eq_NS}). The leading two elements of the Bhattacharyya basis for $\eta_i$ are 
 $$
 v_1(Y_i;\sigma^2,\eta_i)
 =
 \sum_{j=1}^T \frac{Y_{ij}-\eta_i}{\sigma^2},
 \qquad
  v_2(Y_i;\sigma^2,\eta_i)
 =
 -
 \frac{T}{\sigma^2}
 +
\left( \sum_{j=1}^T \frac{Y_{ij}-\eta_i}{\sigma^2} \right)^2.
 $$
We apply Theorem \ref{theo_neyman}. A small calculation yields $A(\sigma^2,\eta_i) = (0, \nicefrac{1}{2T})^\top$ and, after re-arranging, 
$$
u_2^*(Y_i;\sigma^2,\eta_i)
=
\frac{1}{2\sigma^2}
\left(
\frac{\sum_{j=1}^T (Y_{ij}-\overline{Y}_i)^2}{\sigma^2}
-
(T-1)
\right),
$$
which does not depend on $\eta_i$. Summing over the cross-sectional units gives the second-order orthogonalized score equation for $\sigma^2$ as
$$
\sum_{i=1}^N 
u_2^*(Y_i;\sigma^2,\eta_i)
=
\frac{1}{2\sigma^2}
\left(
\frac{\sum_{i=1}^N\sum_{j=1}^T (Y_{ij}-\overline{Y}_i){^2}}{\sigma^2}
-
N(T-1)
\right) = 0,
$$
which yields the degrees-of-freedom corrected estimator $\widehat{\sigma}^2$ in (\ref{eq_sig2_hat}).

Another parameter of interest in this problem is $\mu = \nicefrac{1}{N} \sum_{i=1}^N h(\eta_i)$, for $h$ a known function. This fits our framework, with
$$
u(Y_i; \sigma^2, \eta_i,\mu)
=
h(\eta_i) - \mu.
$$
The $q$-th orthogonalized counterpart to $u$ given by Theorem \ref{theo_neyman} is available in closed form, as
\begin{equation}u_{q}^*(Y_i; \sigma^2, \eta_i,\mu)=h(\eta_i) - \mu+\sum_{k=1}^q \sigma^{k}T^{-\frac{k}{2}}\frac{1}{k!}\nabla_{\eta}^{(k)}h(\eta_i)H_{k}\left(\frac{\sum_{j=1}^T(Y_{ij}-\eta_{i})}{ \sqrt{T}\sigma}\right),\label{eq_ustar_NS}\end{equation}
where $H_{k}$ denotes the $k$-th Hermite polynomial.

Interestingly, (\ref{eq_ustar_NS}) shows that, depending on the form of $h$, the variance of $u_{q}^*$ may converge or diverge as $q$ tends to infinity. Indeed, using a property of Hermite polynomials, the variance is
$$\mbox{Var}\left(u_q^*(Y_i; \sigma^2, \eta_i,\mu)\right)=\sum_{k=1}^q \sigma^{2k}T^{-k}\frac{1}{k!}\left[\nabla_{\eta}^{(k)}h(\eta_i)\right]^2.$$
It is instructive to consider the following three cases: (1) when $h(\eta_i)$ is a polynomial of order $K$ the series is stationary for $q\geq K$; (2) when $h(\eta_i)=\exp(\eta_i)$ the series converges; (3) in contrast, when $h(\eta_i)=\log(\eta_i)$ the series diverges. In the first two cases, using a large $q$ reduces bias without causing variance to diverge, while the third case presents a sharp trade-off between reduced bias and exploding variance as $q$ increases.

\subsection{Nonlinear network regression\label{sec_nonlin}}
Our next example is the nonlinear regression model with $d\geq 1$ outcomes,
\begin{equation}Y_i=m(X_i;\theta_0,\eta_{i0})+\sigma(X_i;\theta_0)\varepsilon_i,\quad \varepsilon_i\,|\, X\sim \mathrm{iid}~{\cal{N}}(0,I_d),\label{eq_network_nonlinreg}\end{equation}
where $m(x;\theta,\eta_i)$ is a $d\times 1$ vector, $\sigma(x;\theta)$ is a $d\times d$ diagonal matrix, and $m$ and $\sigma$ are known functions. We will show below that our CES production function example, in logarithms, fits into this framework.

For this model there are no analytical solutions for the orthogonalized estimators. We thus proceed numerically. To construct Neyman-orthogonal moment functions according to Theorem \ref{theo_neyman} we need to compute $\varSigma_{w_qw_q}(x;\theta,\eta)$, $\varSigma_{w_qu}(x;\theta,\eta,\mu)$, and $b_q(x;\theta,\eta,\mu)$, which involve higher-order derivatives of the conditional likelihood. To compute these derivatives, it is convenient to introduce the operator $\nabla_{m}^q$ that collects all derivatives with respect to 
the $d$-vector $m$ up to order $q$. By the chain rule,
$$\nabla_{\eta_i}^q \ell(y\,|\, x;\theta,\eta_i)=M(x,\theta,\eta_i) \nabla_{m}^q \ell(y\,|\, x;\theta,\eta_i),$$
where the matrix $M$ has an analytical expression given by the multivariate Fa\`a di Bruno formula (\citealp{constantine1996multivariate}).
Given the matrix $M$ it is easy to compute $\varSigma_{w_qw_q}$, $\varSigma_{w_qu}$, and $b_q $. For example,
\begin{align*}
	&\varSigma_{w_qw_q}(x;\theta,\eta_i)\\
	&=M(x,\theta,\eta_i) \,	\mathbb{E}_{\theta,\eta_i}\left(\frac{\nabla_{m}^q \ell(Y_i\,|\, X_i;\theta,\eta_i)}{\ell(Y_i\,|\, X_i;\theta,\eta_i)}\, \frac{\nabla_{m}^q \ell(Y_i\,|\, X_i;\theta,\eta_i)}{\ell(Y_i\,|\, X_i;\theta,\eta_i)}^\top\, \Bigg|\,  X_i=x\right) M(x,\theta,\eta_i)^{\top},
\end{align*}
where the expectation on the right-hand side can be readily computed by relying on formulas for moments of Hermite polynomials. We relegate further details to Appendix \ref{App_sec_implement}. In the next section we present simulations and an empirical application based on a version of (\ref{eq_network_nonlinreg}) designed to study team production.

\paragraph{Example: CES production function (continued).}

Consider the team production model
\begin{equation}Y_j=\beta_0(s_j)\left(\frac{1}{s_j}\sum_{r=1}^{s_j}\eta_{k(j,r)0}^{\gamma_0(s_j)}\right)^{\frac{1}{\gamma_0(s_j)}}\varepsilon_j^{\sigma_0(s_j)},\quad \log \varepsilon_j\,|\, {{\cal{K}}}\sim \mathrm{iid} ~{\cal{N}}\left(0,1\right),\label{eq_teams}\end{equation}
where $s_j$ is the size of team $j=1,...,n$, $(k(j,1),\ldots, k(j,s_j))$ are the $s_j$ workers in team $j$, and the set ${{\cal{K}}}=\{(k(j,1),\ldots, k(j,s_j))\,:\,  j=1,\ldots,n \}$ collects the workers in all teams. Model (\ref{eq_teams}) generalizes Model (\ref{eq_CES_size2}) by allowing for teams of varying sizes. Here we focus on teams of size 1 and 2, as in our application, and impose the normalization $\beta_0(1)=1$. For simplicity we will denote $\beta_0=\beta_0(2)$ and $\gamma_0=\gamma_0(2)$, which are the team size and substitution parameters, respectively, in teams of size 2. 

We now explain how (\ref{eq_teams}) can be written as a special case of (\ref{eq_network_nonlinreg}), for a suitable choice of subsets of observations. To any team $j$ of size 2 involving workers $k$ and $k'$, we associate a team $j_1(j)$ of size 1 only involving worker $k$, and a team $j_2(j)$ of size 1 only involving worker $k'$. This construction results in $N$ subsets of three teams each. We then write the outcomes for these three teams, in logarithms, as
\begin{align}
\log Y_j&=\log \beta_0+\frac{1}{\gamma_0}\log\left(\frac{\eta_{k(j,1)0}^{\gamma_0}+\eta_{k(j,2)0}^{\gamma_0}}{2}\right)+\sigma_0(2)\log \varepsilon_j,\label{eq_subnet_1}\\
\log Y_{j_1(j)}&=\log\eta_{k(j,1)0}+\sigma_0(1)\log \varepsilon_{j_1(j)},\label{eq_subnet_2}\\
\log Y_{j_2(j)}&=\log\eta_{k(j,2)0}+\sigma_0(1)\log \varepsilon_{j_2(j)},\label{eq_subnet_3}
\end{align}
which takes the same form as (\ref{eq_network_nonlinreg}), for $d=3$, $\theta_0=\left(\beta_0,\gamma_0,\sigma_0^2(1),\sigma_0^2(2)\right)^\top$, $Y_i$ the vector of the three outcomes in (\ref{eq_subnet_1})--(\ref{eq_subnet_3}) for subset $i$, and $\eta_{i0}$ the $2\times 1$ vector of worker-specific effects in the corresponding teams.

\begin{remark}{(Implementation in other models)}
In models with discrete outcomes $Y$, one can express the matrices that feature in the expression for $A(x;\theta,\eta,\mu)$ in Theorem \ref{theo_neyman} in closed form, as sums over the support of $Y$. In models with continuous outcomes, one can proceed by simulation as follows, in the spirit of the ``reparameterization trick'' (\citealp{kingma2014auto}). Write $Y=g(X,U;\theta,\eta)$ where $U\,|\, X\sim F_U$ (for example, a standard multivariate Gaussian). Let $U^{(s)}$, $s=1,...,S$, be i.i.d. draws from $F_U$, and let $Y^{(s)}=g(x,U^{(s)};\theta,\eta)$ and $Z^{(s)}=(Y^{(s)},x)$. Assuming that $g$ is a smooth function of $\eta$ one can construct the simulation-based counterpart
\begin{align*}\widehat A(x;\theta,\eta,\mu)&=\left(\sum_{s=1}^Sw_q(Z^{(s)};\theta,\eta)w_q(Z^{(s)};\theta,\eta)^\top\right)^{-1}\\
&\times\left[\sum_{s=1}^Sw_q(Z^{(s)};\theta,\eta)u(Z^{(s)};\theta,\eta,\mu)^\top-\sum_{s=1}^S\nabla_{\eta}^qu\left(g(x,U^{(s)};\theta,\eta),x;\theta,\eta,\mu\right)^\top\right].\end{align*}
When focusing on $\theta_0$ instead of $\mu_0$, one can rely on the simpler expression
\begin{align*}\widehat A(x;\theta,\eta,\mu)=&\left(\sum_{s=1}^Sw_q(Z^{(s)};\theta,\eta)w_q(Z^{(s)};\theta,\eta)^\top\right)^{-1}\sum_{s=1}^Sw_q(Z^{(s)};\theta,\eta)u(Z^{(s)};\theta,\eta)^\top,\end{align*}
and $u_q^*$ can be obtained by regressing $u(Z^{(s)};\theta,\eta)$ on $w_q(Z^{(s)};\theta,\eta)$. We leave the study of the impact of a finite number $S$ of draws on inference about $\mu_0$ and $\theta_0$ to future work. 
\end{remark}

\renewcommand{\theequation}{\arabic{section}.\arabic{equation}}  \setcounter{equation}{0}
\section{Application to team production\label{sec_appli}}

\subsection{Model, data, and implementation}

We wish to estimate the parameters of the team production model in \eqref{eq_subnet_1}--\eqref{eq_subnet_3}. We will be especially interested in estimating the substitution parameter $\gamma$, which drives the nature of complementarities in teams of size 2, and the team size parameter $\beta$, which reflects the premium (or penalty) associated with working together relative to working alone. In addition to estimating production-function parameters, we will also report estimates of a counterfactual random re-allocation of workers to teams. Under random assignment, average output in teams of size 2 can be written as
\begin{equation}\mathbb{E}^{\rm rand}(Y_j)=\frac{2}{n_2(n_2-1)}\sum_{k_1<k_2}{\beta_0}\left[\frac{1}{2}\left({\eta_{k_10}}^{{\gamma_0}}+{\eta_{k_20}}^{{\gamma_0}}\right)\right]^{\frac{1}{{\gamma_0}}}\exp\left(\frac{1}{2}{\sigma_0^2(2)}\right),\label{eq_rand_alloc}
\end{equation}
where $n_2$ denotes the number of authors in teams of size 2. As this quantity is an average over the worker fixed effects, it can be orthogonalized with respect to them using our approach. 

\citet{AhmadpoodJones2019} consider model (\ref{eq_teams}) without the error term $\varepsilon_j$. Here our goal is to address the statistical challenge caused by the presence of a large number of possibly imprecisely estimated fixed effects. An alternative would be to specify a distribution for author heterogeneity conditional on the team network (i.e., for all the $\eta_{i0}$'s conditional on ${{\cal{K}}}$), as in \citet{Bonhomme2021}. An advantage of such a procedure is that, under correct specification, estimates are consistent even in poorly connected networks. This random-effect approach requires, however, to model how authors sort and collaborate in teams. Our approach avoids the need to do so. On the other hand, a fixed-effect approach requires that the author effects can be consistently estimated. In less well-connected networks, the convergence rate will be slower. Orthogonalization to a higher-order allows us to reduce the impact of estimation noise.

We look at the production of academic work in economics. We use data from \citet{DuctorFafchampsGoyalvanderLeij2014}, drawn from the EconLit database. These data contain a large collection of articles, indicated by their ID, together with author identifiers and a measure of {journal quality} proposed by \citet{kodrzycki2006new}. This measure is a ranking between 0 and 100, which we net of multiplicative time effects and will use as our outcome variable. In order to mitigate the variation in author fixed effects over time while ensuring sufficiently many collaborations, we restrict the sample to articles published between 1990 and 1999, written either alone or with a single co-author.\footnote{We have also estimated the model on the entire sample, which ranges from 1970 to 1999. While the estimates of the substitution parameter differ somewhat in this larger sample, they are also less stable due to the fact that connectivity is lower. The estimates of the other parameters are similar to the ones on the 1990-1999 subsample.} We only include authors who produced {at least two sole-authored articles} during the sampling period. 

Our sample contains 91,626 articles, 10\% of which are co-authored, and 16,408 authors. Average journal quality differs greatly across authors, with the 10th percentile of the quality measure being 0.4, the median being 0.9, and the 90th percentile being 8.5. The between-author variance in journal quality is 42\% of the overall variance. The distribution of journal quality, in turn, is skewed to the right, with a median of 0.6, a 90th percentile of 12, and a 99th percentile of 52. The number of publications per author varies substantially, with a 10th percentile of 2, a median of 4, and a 90th percentile of 13.

To implement our approach, we construct subsets of three papers, one co-authored ($j$) and two sole-authored ($j_1(j),j_2(j)$), as described in (\ref{eq_subnet_1})--(\ref{eq_subnet_3}). The score for $\theta$ based on subset $i$ then involves the three teams $j$, $j_1(j)$, and $j_2(j)$. Proceeding in this way is helpful as it limits the dimension of the parameter $\eta_i$ to two. This is not only in line with the assumptions we make in deriving asymptotics, but also helpful in terms of computation. Moreover, it reduces the number of derivatives that need to be computed. The number of derivatives nevertheless remains substantial, as we need to compute $9$ derivatives at order 2, $19$ at order 3, and $55$ at order 5, for example; these are the derivatives with respect to the three-dimensional mean vector $m$ in (\ref{eq_network_nonlinreg}), from which the derivatives with respect to $\eta_i$ are obtained by the chain rule. Yet, using the computational remarks from Subsection \ref{sec_nonlin}, this remains practical.

Finally, we exploit the network structure of the data to perform our sample splitting. For every worker, we construct a preliminary estimator of her fixed effect (in logs) as the average quality of her single-authored papers, except for one that we select at random and use later in estimation. This strategy is feasible due to our sample restriction. For each subset $i$ of three teams, we then stack the two worker fixed effects together to form our preliminary estimate $ \widehat{\eta}_i$. We next estimate the parameters $\beta_0,\gamma_0,\sigma_0^2(1),\sigma_0^2(2)$ on the sample from which all these single-authored articles have been removed. In the present case, $\widetilde{u}$ in (\ref{thetaEstimation}) has four components that correspond to the score with respect to all the parameters, and the weight matrix $\widetilde{W}$ is irrelevant since the problem is just-identified. In order to limit the variability due to the choice of split, we average parameter estimates across 100 random splits, through cross-fitting. The bias in the parameter estimates takes a complex form due to the team network environment. In Appendix \ref{AppC} we assess the ability of our orthogonalization approach to alleviate this bias in a Monte Carlo simulation.

\subsection{Empirical estimates}

Table \ref{tab_appli} shows the estimates of $\beta_0$, $\gamma_0$, $\sigma_0^2(2)$, and $\sigma_0^2(1)$ for various estimators. These are the plug-in estimator based on the preliminary estimates $\widehat{\eta}_i$ and six estimators based on Neyman-orthogonalized moments, for $1\leq q \leq 6$. In addition to point estimates, we report robust standard errors based on the parametric bootstrap.\footnote{Bootstrap replications are based on Neyman-orthogonalized estimates of $\beta_0$, $\gamma_0$, $\sigma_0^2(2)$, and $\sigma_0^2(1)$ to order $q=6$, together with the sample-split estimates $\widehat{\eta}_i$ of author effects. Within each bootstrap replication, we cross-fit the estimates 10 times. Results are based on 200 bootstrap replications. We report robust estimates of bootstrap standard deviations, obtained as interdecile ranges of bootstrap draws divided by $2\Phi^{-1}(0.90)=2.5631$ (\citealp{machado2005bootstrap}). While using robust estimates has little impact on standard errors of parameter estimates in Table \ref{tab_appli}, this deals with the presence of a few influential bootstrap draws in the calculation of re-allocation effects in Table \ref{tab_appli2}.}

\begin{table}
	\caption{Estimation results\label{tab_appli}}
	\begin{center} 
	\begin{tabular}{|c||cccc|c|}\hline\hline 
	& Substitution ${\gamma}$ & Team size ${\beta}$ & Variance ${\sigma^2(2)}$ & Variance ${\sigma^2(1)}$ & Diagnostic \\ \hline\hline 
	Plug-in & $\underset{\small (0.0386)}{0.1267}$ & $\underset{\small (0.0222)}{1.2893}$ & $\underset{\small (0.0218)}{1.6232}$ & $\underset{\small (0.0274)}{1.6395}$ & - \\ 
	$q=1$ & $\underset{\small (0.1758)}{-1.8905}$ & $\underset{\small (0.0269)}{1.3260}$ & $\underset{\small (0.0255)}{1.6793}$ & $\underset{\small (0.0247)}{1.7330}$ &  $592.65$\\ 
	$q=2$ & $\underset{\small (0.1939)}{0.7678}$ & $\underset{\small (0.0305)}{1.2956}$ & $\underset{\small (0.0241)}{1.4339}$ & $\underset{\small (0.0262)}{1.4683}$ &$195.77$\\ 
	$q=3$ & $\underset{\small (0.1941)}{0.4702}$ & $\underset{\small (0.0307)}{1.2845}$ & $\underset{\small (0.0238)}{1.4399}$ & $\underset{\small (0.0259)}{1.4407}$ &$176.48$\\ 
	$q=4$ & $\underset{\small (0.1873)}{0.4176}$ & $\underset{\small (0.0309)}{1.2838}$ & $\underset{\small (0.0239)}{1.4348}$ & $\underset{\small (0.0253)}{1.4214}$ & $50.770$ \\ 
	$q=5$ & $\underset{\small (0.1798)}{0.4143}$ & $\underset{\small (0.0309)}{1.2836}$ & $\underset{\small (0.0239)}{1.4327}$ & $\underset{\small (0.0244)}{1.4180}$ & $43.708$ \\ 
	$q=6$ & $\underset{\small (0.1798)}{0.4145}$ & $\underset{\small (0.0317)}{1.2836}$ & $\underset{\small (0.0239)}{1.4315}$ & $\underset{\small (0.0242)}{1.4165}$ & - \\ \hline\hline 
	\end{tabular} 
	\end{center}
	
\raggedright	
\par\textit{{\footnotesize Notes: Point estimates based on $q$-ordered orthogonalized estimators, cross-fitted estimates (100 splits). Robust parametric bootstrap standard errors in parentheses (200 replications). In the last column we report the diagnostic for the orthogonality order $q$ (see Appendix \ref{App_q}, bootstrapped with 200 replications), which is asymptotically $\chi^2(4)$ if bias at order $q$ is statistically negligible.}}
\end{table}

Starting with the substitution parameter $\gamma$, the uncorrected estimate is $0.13$, which is close to the Cobb-Douglas case. The value of the first-order Neyman-orthogonalized estimate is quite different. However, since the preliminary estimates of the author fixed effects are based on very few observations, we do not expect this estimator to adequately correct for bias. This is confirmed by the fact that all other Neyman-orthogonalized estimates, for $q\in\{2,\ldots,6\}$, range between $0.41$ and $0.77$, which is higher than the plug-in estimate, and very different from the first-order orthogonalized estimate. Relative to the plug-in, the orthogonalized estimates with $q\geq 2$ all indicate somewhat less complementarity between authors in team production. Notice the stability of estimates for larger values of $q$. A substitution parameter $\gamma=0.4$ corresponds to a production function that is supermodular in worker inputs; see Figure \ref{fig_prodf} in Appendix \ref{App_fig} for a graphical illustration.

Turning to the other parameters, the estimates of the team size parameter $\beta$ are virtually unaffected by the orthogonalization. This suggests the bias is limited for this parameter. Its value is close to $1.3$, implying that producing a paper with a co-author increases the paper's quality to some extent. Next, the log-error variance $\sigma^2(2)$ in teams of two co-authors is larger when using plug-in estimates ($1.6$) than when using orthogonalization with $q\geq 2$ ($1.4$), suggesting that the plug-in and first-order corrected estimates are biased upward. Lastly, the variance $\sigma^2(1)$ in teams of a single author is also larger under the plug-in estimator.

An interesting feature of Table \ref{tab_appli} is that point-estimates and standard errors appear to converge as $q$ increases. This suggests the absence of a sharp bias-variance trade-off as a function of $q$. As we have seen in the case of the Neyman-Scott example in Subsection \ref{subsec_panel}, this phenomenon is specific to the model and target parameter of interest, and the variance may converge or diverge as $q$ grows depending on the context.

One potential explanation for convergence in the present setting is that Model (\ref{eq_teams}) implies some non-trivial restrictions on the parameters $\gamma,\beta,\sigma^2(1),\sigma^2(2)$ that do not depend on the author-specific effects $\eta_i$, as we show in Appendix \ref{AppD}. Moment functions that are independent of $\eta$ are common in panel data settings, obtained e.g. by differencing, quasi-differencing, or functional differencing. In some models, such as models with discrete outcomes, such functions may not exist. In Appendix \ref{AppD} we exploit two types of restrictions as robustness checks. Our findings suggest that, while those restrictions seem broadly consistent with the higher-order orthogonal estimates reported in Table \ref{tab_appli}, using them directly for estimation may lead to very imprecise estimates. In contrast, Table \ref{tab_appli} suggests that higher-order Neyman-orthogonality may be a successful approach to approximate such functions in settings where those exist. 

In the last column of Table \ref{tab_appli} we report our diagnostic statistic for the orthogonality order $q$, described in Appendix \ref{App_q}.\footnote{The statistic in Appendix \ref{App_q} depends on a variance matrix estimate $\widehat V$, which we compute based on the parametric bootstrap, cross-fitted 10 times (200 replications).} A value larger than the 95-th quantile of the $\chi^2(4)$ distribution (9.488) should be interpreted as evidence that the order-$q$ and order-$(q+1)$ orthogonal moments differ. The values of the statistic reported in the table suggest that biases remain statistically detectable even at  $q=5$. This is likely due to the large sample size, since point estimates are economically similar between $q=4$ and $q=6$.

\begin{table}[tbp]
	\caption{Empirical estimates: average output\label{tab_appli2}}
	\begin{center} 
	\begin{tabular}{|c||cc|}\hline\hline 
	& Average output & Counterfactual\\\hline\hline 
	Plug-in & $\underset{\small (0.2038)}{8.4434}$ & $\underset{\small (0.1676)}{7.2045}$ \\ 
	$q=1$ & $\underset{\small (0.5120)}{6.8366}$ & $\underset{\small (0.4260)}{5.4330}$ \\ 
	$q=2$ & $\underset{\small (0.5482)}{9.0903}$ & $\underset{\small (0.7685)}{8.6758}$ \\ 
	$q=3$ & $\underset{\small (0.3778)}{6.1290}$ & $\underset{\small (0.3617)}{5.5553}$ \\ 
	$q=4$ & $\underset{\small (0.3995)}{7.4412}$ & $\underset{\small (0.4399)}{6.6691}$ \\ 
	$q=5$ & $\underset{\small (0.2380)}{6.9758}$ & $\underset{\small (0.2985)}{6.2593}$ \\ 
	$q=6$ & $\underset{\small (0.2640)}{7.1168}$ & $\underset{\small (0.3449)}{6.3805}$ \\
    \hline\hline 
	\end{tabular}
	\end{center}
	
	\raggedright
    \par\textit{{\footnotesize Notes: Average output (the value in the data among co-authored articles is $6.9995$, robust standard error $0.2231$), and counterfactual average output in a random allocation. Point estimates based on orthogonalized estimators to order $q$, cross-fitted estimates (100 splits). Robust parametric bootstrap standard errors in parentheses (200 replications).}}
\end{table}

Lastly, we report estimates of average journal quality in a counterfactual scenario where authors are randomly assigned across teams of two co-authors, see (\ref{eq_rand_alloc}). The first column in Table \ref{tab_appli2} shows estimates of the average output in the empirical allocation. This quantity can be estimated without bias as the sample mean of the journal quality variable among co-authored articles, which is equal to $7.0$. We see that the plug-in estimate is $8.4$, larger than the empirical value. In comparison, Neyman-orthogonalized estimates for $q\geq 3$ range between $6.1$ and $7.4$, and estimates for $q=5$ and $q=6$ are closest to the empirical value. The second column in Table \ref{tab_appli2} shows estimates of average article quality under random assignment of authors to teams, using the plug-in method and Neyman-orthogonalized estimates to order $q\geq 1$.\footnote{To speed up computation, we approximate (\ref{eq_rand_alloc}) using a random subset of 5000 pairs of authors, for each random sample split (and each bootstrap replication).} The estimates vary with the order of orthogonalization. When taking $q\geq 4$, estimates range between $6.2$ and $6.7$. In addition, comparing the two columns of Table \ref{tab_appli2} shows that, irrespective of the order of orthogonalization, the estimates of average output are lower in the counterfactual scenario where workers are randomly allocated across teams.  

The main takeaway from Table \ref{tab_appli2} is that randomly allocating authors among teams would tend to lower average paper quality. This is due to two economic forces. The first one is complementarity in production, as reflected by estimates of $\gamma$ lower than $1$. The second force is positive sorting. Indeed, the preliminary estimates of worker fixed effects are positively correlated within teams in the data. In the presence of complementarity, decreasing assortative matching leads to lower output, which is what we find in Table \ref{tab_appli2}.

\renewcommand{\theequation}{\arabic{section}.\arabic{equation}}  \setcounter{equation}{0}

\section{Asymptotic properties\label{sec_asympt}}

In this section, we show that, under higher-order orthogonality,
the estimators $\widehat{\theta}$ and $\widehat{\mu}$
introduced in Subsection~\ref{subsect:Estimation} are $\sqrt{n}$-consistent and asymptotically normal
under appropriate assumptions, even if the convergence rate of $\widehat{\eta}_i$ is slower than $\sqrt{n}$. We focus on deriving the asymptotic distribution of $\widehat \mu$, assuming that we have already  worked out the corresponding asymptotic result of $\widehat \theta$. However, the corresponding theory for $\widehat \theta$ is actually a special case of our results for $\widehat \mu$, where $\theta$ is dropped from the arguments, $\mu$ is replaced by $\theta$, and $u$ is replaced by $\widetilde u$. Thus, our focus on $\widehat \mu$ is without loss of generality.

\subsection{Notation}

For the presentation of the asymptotic theory, it is useful to be explicit about which parameters depend on the sample size and which ones do not. Recall that $n$ is the total number of observations in $(Z_1,...,Z_N)$, where each $Z_i$ comprises $n_i$ observations. In the asymptotic sequence, we let $N$ and $n_i$ depend on $n$, although we do not explicitly indicate this dependence. For example, in a panel data model, our assumptions allow both $N$ and $T$ to grow as the number $NT$ of observations tends to infinity. 

To indicate the dependence on the sample size, we will write $\eta_n$ and $\mu_n$ instead of $\eta$ and $\mu$ in this section. While the dimension of $\mu$ is not changing with $n$, the true parameter $\mu_{0,n}$  is implicitly defined as the solution 
of $\sum_{i=1}^N\mathbb{E}_{\theta_0,\eta_{0,n}}\left( u(Z_i; \theta_0, \eta_{0,n,i},\mu)\right) = 0$, which may depend on $n$. By contrast, the parameter $\theta$ and its true value $\theta_0$ are independent of $n$. 

Remember also that $n=\sum_{i=1}^N n_i$, and note that if the observations within each unit $i$
are independent, then we have $\ell(y_i\,|\, x_i;\theta,\eta_{n,i}) = \prod_{j=1}^{n_i} 
\ell(y_{ij}\,|\, x_{ij};\theta,\eta_{n,i})$. Hence, in the case of the score for $\theta$,
\begin{align*}
    u(Z_i; \theta, \eta_{n,i}) = 
    \sum_{j=1}^{n_i} \frac{\partial \log\ell(y_{ij}\,|\, x_{ij};\theta,\eta_{n,i})}{\partial\theta}.
\end{align*}
More generally, whenever $n_i \to \infty$ we expect that $u$ scales linearly
with $n_i$, {explaining the scaling of $u(Z_i; \theta, \eta_{n,i})$ and of various other terms in Assumption~\ref{ass:asymptotic1} below.}

Finally, in what follows, $D^m_{\eta_{n,i}}$ denotes the derivative operator with respect to $\eta_{n,i}$, and
${\mathbb K}_{q,n,i} = \{ m \in \mathbb{N}^{{\rm dim}(\eta_{n,i})} : 1\leq \sum_{r} m_r \leq q \}$.

\subsection{A useful lemma}

With this notation in hand, we now state our first assumption.

\begin{assumption}
    \label{ass:asymptotic1}
     \phantom{a}
    \begin{enumerate}[(i)]
         \item   We have 
    $\left[\frac 1 n \sum_{i=1}^N  
    \frac{\partial u^\top (Z_i;  \widehat \theta,\widehat \eta_{n,i},\widehat\mu_n  )} {\partial \mu}
     \right] W \left[\frac 1 {\sqrt{n}} \sum_{i=1}^N  u(Z_i;  \widehat \theta,\widehat \eta_{n,i},\widehat\mu_n  ) \right]  =  o_P(1)$, for some non-random symmetric 
           positive definite weight matrix $W$.

        \item As $n\rightarrow \infty$, $(\widehat \theta,\widehat \eta_n ,  \widehat \mu_n)$ is contained
        in a convex neighborhood ${\cal B}_n$ of $(\theta_0,\eta_{0,n},\mu_{0,n})$. Let ${\cal B}_{n,i}$ be the convex neighborhood of $(\theta_0,\eta_{0,n,i},\mu_{0,n})$ obtained by intersecting ${\cal B}_n$ with the parameter subspace for observation $i$.
        
        \item 
       $\max_i \dim(\eta_{n,i})=O(1)$. 
        
        \item For every  $i$, the function  $u(Z_i,\theta,\eta_{n,i},\mu)$ is   $(q+1)$ times continuously differentiable 
         in the parameters $(\theta,\eta_{n,i},\mu)$, and we assume that for all its components all the partial derivatives of $u(Z_i;\theta,\eta_{n,i},\mu)$ up to order $(q+1)$ are  bounded in 
         absolute value
         by  $n_iC_{n,i}(Z_i) \geq 0$, uniformly in the neighborhood ${\cal B}_{n,i}$, such that
         $\frac{1}{n} \sum_{i=1}^N  n_i\mathbb{E} \left[ C_{n,i}(Z_i)^2 \right] = O(1)$. In addition, for all $m \in {\mathbb K}_{q,n,i}$ we have
$\mathbb{E}\big[ \| D^{m}_{\eta_{n,i}} u(Z_i;\theta_0,\eta_{0,n,i},\mu_{0,n}) \|^2 \big] \leq n_i \widetilde{C}_{n,i}$
for non-random $\widetilde{C}_{n,i} \geq 0$ such that
$\frac{1}{n} \sum_{i=1}^N n_i \widetilde{C}_{n,i}^2 = O(1)$.

         \item 
        $\widehat \mu_n - \mu_{0,n} = o_P(1)$ and $\frac 1 n \sum_{i=1}^N n_i \mathbb{E}\left(  \left\|
            \widehat \eta_{n,i}-\eta_{0,n,i} \right\|^{2(q+1)}\right) = o(n^{-1})$.

        \item   $\widehat \theta = \theta_0+\frac 1 n \sum_{i=1}^N  \psi_{n,i} + o_P(n^{-1/2})$, where $\psi_{n,i}$ is a function of $Z_i$ with $\mathbb{E}   (\psi_{n,i}) = 0$ and $\frac{1}{n} \sum_{i=1}^N \mathbb{E}  \left(\left\| \psi_{n,i} \right\|^2 \right)=O(1)$.

          \item The probability limits
           $$G_\mu = \plim_{n \rightarrow \infty}  \frac 1 {n} \sum_{i=1}^N \frac{\partial u(Z_i;   \theta_0, \eta_{0,n,i} ,  \mu_{0,n})}{\partial \mu^{\top}},\quad G_\theta = \plim_{n \rightarrow \infty}  \frac 1 {n} \sum_{i=1}^N \frac{\partial u(Z_i;   \theta_0, \eta_{0,n,i} ,  \mu_{0,n})}{\partial \theta^{\top}}$$
           exist, and ${\rm rank}(G_\mu)={\rm dim}(\mu)$.

    \end{enumerate}
\end{assumption}

Part $(i)$ in Assumption \ref{ass:asymptotic1} is satisfied if $\widehat\mu_n$ is computed using GMM, see (\ref{muEstimation}). In Part $(ii)$, the neighborhood ${\cal B}_n$ depends on the sample size $n$, partly because the number of nuisance parameters of $\eta_{n,i}$ generally depends on $n$. Part $(iii)$ assumes that the maximal dimension of $ \eta_{n,i}$ is bounded as $n \rightarrow \infty$. Part $(iv)$ requires the derivatives of the moment functions (properly rescaled) to be suitably bounded. Its second requirement bounds the second moment of those derivatives by $n_i$, rather than by $n_i^2$ as the first bound would deliver. Under Assumption \ref{ass:asymptotic2}$(i)$ these derivatives have mean zero, so that their second moment is a variance; when $u(Z_i;\cdot)$ is a sum over the $n_i$ observations in $Z_i$, as in our panel data and network examples, this variance is of order $n_i$. The first half of Part $(v)$ is a high-level
consistency assumption for $\widehat \mu_n$, which
can be justified by guaranteeing that the
objective function in \eqref{muEstimation} converges
uniformly to a population counterpart that has a 
unique minimum at $\mu_0$. The second half of Part $(v)$ is the rate requirement on the preliminary estimates $\widehat{\eta}_{n,i}$, imposing a rate faster than $n^{-\nicefrac{1}{2(q+1)}}$. Part $(vi)$ requires $\widehat{\theta}$ to be asymptotically linear, in particular requiring $  \widehat \theta - \theta_0  = O_P(n^{-1/2})$. In the case where $\mu_{0,n}=\theta_0$ this condition is not needed. Lastly, Part $(vii)$ assumes existence of Jacobian matrices and a rank condition.

\begin{lemma}
    \label{lemma:Expansion}
     Under  Assumption~\ref{ass:asymptotic1}  we have
     \begin{align*}
             &\sqrt{n} \left( \widehat \mu_n - \mu_{0,n} \right)
             \\
             & \; \; =  - \left( G^{\top}_{\mu} \, W \, G_\mu  \right)^{-1} \, G^{\top}_{\mu}  \,W  \left\{ \frac 1 {\sqrt{n}} \sum_{i=1}^N 
             \Big[ u(Z_i;   \theta_0, \eta_{0,n,i} ,  \mu_{0,n}) +  G_{\theta} \, \psi_{n,i} \Big]
                   + R_n
             \right\} + o_P(1),
     \end{align*}
     where   
     \begin{align*}
          R_{n} =  \frac 1 {\sqrt{n}} \sum_{i=1}^N \sum_{m \in {\mathbb K}_{q,n,i}} \frac{1}{m!} \left[ D^m_{\eta_{n,i}} u(Z_i;   \theta_0, \eta_{0,n,i} ,  \mu_{0,n}) \right]  \left(\widehat \eta_{n,i} - \eta_{0,n,i}\right)^{m}.
     \end{align*}
\end{lemma}

\subsection{Main result}

We are now in position to establish the main result of this section, which concerns root-$n$ consistency and asymptotic normality of estimators based on orthogonal equations. For this, we first state our second assumption.

\begin{assumption}
    \label{ass:asymptotic2}
     \phantom{a}
    \begin{enumerate}[(i)]
         \item The moment function 
         $u(Z_i;  \theta, \eta_{n,i} ,  \mu)$ is Neyman-orthogonal to order $q$, and furthermore
          $\sum_{i=1}^N\mathbb{E} \left( u(Z_i;  \theta_0, \eta_{0,n,i} ,  \mu_{0,n})\right)=0$.
      
     \item
     $\widehat \eta_{n,i}$ are independent of $(Z_1,\ldots,Z_N)$ for all $i$.
     
     \item 
     The $Z_1,\ldots,Z_N$ are independent  across $i$.    
     
     \item 
        $\xi_{n,i}= u(Z_i;   \theta_0, \eta_{0,n,i} ,  \mu_{0,n}) + G_{\theta} \, \psi_{n,i}$ satisfies Lindeberg's condition,\footnote{
    Formally (for scalar $\xi_{n,i}$): for any $\epsilon > 0$, $\frac{1}{s_n^2} \sum_{i=1}^N \mathbb{E}\left[ \xi_{n,i}^2 \cdot \mathbbm{1}(|\xi_{n,i}| > \epsilon s_n)\right] \to 0$ as $n \to \infty$, where $s_n^2 = \sum_{i=1}^N \mathrm{Var}(\xi_{n,i})$ and $\mathbbm{1}$ is the indicator function. Whenever $\mathbb{E}(\xi_{n,i}) \neq 0$, the condition is imposed on the centered variables $\xi_{n,i}-\mathbb{E}(\xi_{n,i})$; note that $\sum_{i=1}^N \mathbb{E}(\xi_{n,i})=0$. When $\xi_{n,i}$ is a vector, we impose this condition on $\lambda^\top \xi_{n,i}$ for every non-zero conformable vector $\lambda$, and appeal to the Cram\'er-Wold device.}
    and the following probability limit exists:
     $$
        V_\xi = \plim_{n \rightarrow \infty} \frac 1 {n} \sum_{i=1}^N  {\rm Var}\left( \xi_{n,i} \right) .
     $$

   \end{enumerate}
\end{assumption}

Part $(i)$ in Assumption \ref{ass:asymptotic2} requires $u$ to be Neyman-orthogonal in the sense of Definition \ref{def_ortho}. Part $(ii)$ requires the preliminary estimates to be independent from the estimation sample. With independent observations, this can be achieved by sample splitting. Part $(iii)$ imposes independence between the $Z_i$'s. We impose this assumption to simplify the presentation. It is straightforward to modify the variance expression in Theorem \ref{th:Asymptotic} below to account for particular forms of dependence (e.g., clustered) by using an appropriate expression for the matrix $V_{\xi}$ introduced in Part $(iv)$.  

The following theorem provides an asymptotic characterization of $\widehat{\mu}_{n}$. 

\begin{theorem}
    \label{th:Asymptotic}
     Let  Assumptions~\ref{ass:asymptotic1} and \ref{ass:asymptotic2} hold with the same value of $q \in \{1,2,3,\ldots\}$. Then we have
     \begin{align*}
             \sqrt{n} \left( \widehat \mu_n - \mu_{0,n} \right) &\overset{d}{\rightarrow} {\cal N}\big(0,   \,   \left( G^{\top}_{\mu} \, W \, G_\mu  \right)^{-1} 
             G^{\top}_{\mu}  \,W   \, V_\xi \, W G_\mu   \left( G^{\top}_{\mu} \, W \, G_\mu  \right)^{-1}  \big) .
     \end{align*}
\end{theorem}

Note that, although we leave the dependence on $q$ implicit in Theorem \ref{th:Asymptotic}, the asymptotic variance does depend on the order of orthogonality $q$ that the moment function satisfies. Note also that $ \widehat \mu_n$ in the theorem is based on a single set of preliminary estimates $\widehat{\eta}_i$. The variability caused by the use of a single sample split can be mitigated through the use of cross-fitting, as we do in the application.

\section{Final remarks\label{sec_final}}

In this paper we show how to construct higher-order Neyman-orthogonal moment functions in conditional-likelihood models. We use these functions, together with sample splitting, to reduce bias in estimation. Our application suggests that our higher-order corrections can be effective in network settings with fixed effects. An area of application is to double/debiased machine learning with fixed effects, where the nuisance parameters contain some components, such as low-dimensional functions, for which first-order orthogonality may suffice. However, for such applications it is important to extend the approach to non-likelihood models. As $q$ increases, orthogonalization imposes growing demands on the likelihood structure -- to achieve first-order orthogonality it is sufficient for the score to have mean zero, while to achieve second-order orthogonality our approach requires the information identity to hold, for example. {This reflects a trade-off between the robustness to nuisance parameters that our method achieves and robustness to model misspecification.} In follow-up work (\citealp{bonhomme2026higher}), we propose a construction of orthogonal functions in semi-parametric models defined by moment conditions. In the absence of a conditional-likelihood structure, our approach exploits independence between observations to achieve higher-order orthogonality.

\clearpage

\setlength{\bibsep}{0pt}

\clearpage

\appendix

\begin{center}
	{ {\LARGE APPENDIX} }
\end{center}

\renewcommand{\theequation}{\thesection.\arabic{equation}}  \setcounter{equation}{0}

\section{Proofs\label{App_proofs}}

\subsection{Proof of Theorem \ref{theo_neyman}}

Before proving the theorem, it is useful to establish the following lemma, which we prove in Appendix \ref{App_prooflemma}.

\begin{lemma}\label{lemma:helpful}
Let $q \in \{1,2,3,\ldots\}$, and let $x$ be some realization of the covariates.
Remember that $\nabla^q_\eta$ and 
$w_q(z;\theta,\eta) $ 
are vectors of dimension $k_q = \sum_{p=1}^q d_p$, and 
that
$\varSigma_{w_q w_q}(x;\theta,\eta)$ is a $k_q  \times k_q$ matrix. 
We assume that  $\varSigma_{w_qw_q}(x;\theta,\eta)$ is invertible, and we define
$$
\widetilde{w}_q(z;\theta,\eta) = \varSigma_{w_qw_q}(x;\theta,\eta)^{-1} w_q(z;\theta,\eta).
$$
Then, 
\begin{align}
\mathbb{E}_{\theta,\eta} \left[ (\nabla^q_\eta)^\top \,  \widetilde{w}_q(Z;\theta,\eta)  \big| X=x \right] =  
\left( \begin{array}{ccccc} 
         - \mathbb{I}_{d_1}  & 0 & 0 & \cdots & 0 \\
            0 & +   \mathbb{I}_{d_2} & 0 & \cdots & 0 \\
            0 & 0 & -     \mathbb{I}_{d_3} & \cdots & 0 \\
            \vdots &  \vdots &   \vdots & \ddots & \vdots \\
            0 & 0 & 0 & \cdots & (-1)^q  \;   \mathbb{I}_{d_q} 
            \end{array} \right) ,
   \label{lemma:helpful:result}         
\end{align}
where the diagonal $k_q  \times k_q$ matrix on the right hand side is obtained by
stacking $ (-1)^p \;  \mathbb{I}_{d_p}  $ on the diagonal for $p = 1,\ldots,q$,
and $\mathbb{I}_{d_p} $ is the identity matrix of dimensions $d_p$.
\end{lemma}

\begin{proof}[\bf Proof of Theorem~\ref{theo_neyman}]
Define  
$c_q(x;\theta,\eta,\mu) = \left[ \varSigma_{w_q u}(x;\theta,\eta,\mu) - b_q(x;\theta,\eta,\mu) \right]^\top$.
In the proof of Lemma~\ref{lemma:helpful} in Appendix \ref{App_prooflemma} we introduce the notation $ \widetilde{w}_q^{\,m} (z;\theta,\eta)$, $m \in \mathcal{C}_q$, for the elements of the 
$k_q$-vector $ \widetilde{w}_q(z;\theta,\eta)$. Analogously, we now use $c^{m}_q(x;\theta,\eta,\mu) $ to denote the columns of the $({\dim u}) \times k_q$-matrix
$c_q(x;\theta,\eta,\mu) $, that is,
$c^{m}_q(x;\theta,\eta,\mu) $ is a $({\dim u})$-vector
for every $m \in \mathcal{C}_q$.
We have
\begin{align}
     c^{m}_q(x;\theta,\eta,\mu) &=
      \mathbb{E}_{\theta,\eta}\left[  \frac{D^m_\eta  \ell(y\,|\,  x;\theta,\eta)}{ \ell(y\,|\,  x;\theta,\eta)}  u(Z;\theta,\eta,\mu) \, \Bigg|\,  X=x\right]
      - D^m_\eta \, \mathbb{E}_{\theta,\eta}(u(Z;\theta,\eta,\mu) \vert X=x) 
  \nonumber  \\
     &=  \int   [D^m_\eta  \ell(y\,|\,  x;\theta,\eta)] \, u(z;\theta,\eta,\mu)   dy 
     -   D^m_\eta \  \int  \ell(y\,|\,  x;\theta,\eta) \,  u(z;\theta,\eta,\mu)  dy
  \nonumber \\
       &=     -  \sum_{S \subsetneq \{1,\ldots,|m|\}}   \int   [D^{m_S}_\eta \, \ell(y\,|\,  x;\theta,\eta)] [ D^{m_{-S}}_\eta \, u(z;\theta,\eta,\mu)]  dy ,
     \label{ExpressionForTermCm}  
\end{align}
where $z=(y,x)$ and in the last step we applied the product rule for differentiation
as in \eqref{ApplyProductRule} below, but the term for $S=  \{1,\ldots,|m|\}$ cancels with the term that stems from $ \varSigma_{w_q u}(x;\theta,\eta,\mu)$,
which explains why we only sum over subsets $S$ that are different from $\{1,\ldots,|m|\}$.

Next, by the definition of $u_q^*$ and $ A(x;\theta,\eta,\mu)$, we have
\begin{align*}
u_q^*(z;\theta,\eta,\mu) &= u(z;\theta,\eta,\mu) - A(x;\theta,\eta,\mu)^\top w_q(z;\theta,\eta) \\
&= u(z;\theta,\eta,\mu) - \left[ \varSigma_{w_q u}(x;\theta,\eta,\mu) -  b_q(x;\theta,\eta,\mu)  \right]^\top  \widetilde{w}_q(z;\theta,\eta)
\\
&=  u(z;\theta,\eta,\mu)      - \sum_{v \in \mathcal{C}_q} c^{v}_q(x;\theta,\eta,\mu)   \,  \widetilde{w}_q^{\,v} (z;\theta,\eta) .
\end{align*}
Let $m \in \mathcal{C}_q$.  Applying the operator $D^m_\eta$ to the last equation and again using the product rule for differentiation in the same way as before, we find
\begin{align*}
 D^m_\eta u_q^*(z;\theta,\eta,\mu) 
&=  D^m_\eta u(z;\theta,\eta,\mu)   - 
 \sum_{S \subseteq \{1,\ldots,|m|\}}  \sum_{v \in \mathcal{C}_q} \left[D^{m_{-S}}_\eta \, c^{v}_q(x;\theta,\eta,\mu) \right]   \left[D^{m_S}_\eta \, \widetilde{w}_q^{\,v} (z;\theta,\eta) \right]  .
\end{align*}
Applying the conditional expectation operator to this and using Lemma~\ref{lemma:helpful} we obtain
\begin{align*}
& \mathbb{E}_{\theta,\eta}\left[  D^m_\eta u_q^*(Z;\theta,\eta,\mu)  \big| X=x \right] 
\\
&=  \mathbb{E}_{\theta,\eta}\left[ D^m_\eta u(Z;\theta,\eta,\mu)  \big| X=x \right]   
\\ & \quad
- 
 \sum_{S \subseteq \{1,\ldots,|m|\}}  \sum_{v \in \mathcal{C}_q} \left[D^{m_{-S}}_\eta \, c^{v}_q(x;\theta,\eta,\mu) \right] 
     \underbrace{\mathbb{E}_{\theta,\eta}\left[D^{m_S}_\eta \, \widetilde{w}_q^{\,v} (Z;\theta,\eta)   \big| X=x  \right]
       }_{= (-1)^{|S|} \, \mathbbm{1}\left\{ m_S = v \right\}}  
     \\
&=  \mathbb{E}_{\theta,\eta}\left[  D^m_\eta u(Z;\theta,\eta,\mu)  \big| X=x \right]   -
 \sum_{\emptyset \neq S \subseteq \{1,\ldots,|m|\}}    (-1)^{|S|}   \left[D^{m_{-S}}_\eta \, c^{m_S}_q(x;\theta,\eta,\mu) \right],
\end{align*}
where in the last step we used that for $S = \emptyset$
the indicator $\mathbbm{1}\left\{ m_S = v \right\}$
is always zero (because $v \in \mathcal{C}_q$ never has 
length zero), but for $S \neq \emptyset$ there is always
exactly one $v \in \mathcal{C}_q$ that satisfies $m_S=v$,
that is, in that second case we just remove the sum
over $v$ and replace $v$ by $m_S$ throughout.
Next,
plugging in the expression for $c^{m}_q(x;\theta,\eta,\mu)$ in equation \eqref{ExpressionForTermCm} above we find
\begin{align*}
& \mathbb{E}_{\theta,\eta}\left[  D^m_\eta u_q^*(Z;\theta,\eta,\mu)  \big| X=x \right] 
\\
&=  \mathbb{E}_{\theta,\eta}\left[  D^m_\eta u(Z;\theta,\eta,\mu)  \big| X=x \right] 
    \\ & \quad
  +
 \sum_{\emptyset \neq  S \subseteq \{1,\ldots,|m|\}}    (-1)^{|S|}   D^{m_{-S}}_\eta \left( 
      \sum_{T \subsetneq S}   \int   [D^{m_T}_\eta \, \ell(y\,|\,  x;\theta,\eta)] [ D^{m_{S \setminus T}}_\eta \, u(z;\theta,\eta,\mu)]  dy
     \right).
\end{align*}
By again using the product rule for differentiation to apply $D^{m_{-S}}_\eta $ to the product in the last term, we obtain
\begin{align}
& \mathbb{E}_{\theta,\eta}\left[  D^m_\eta u_q^*(Z;\theta,\eta,\mu)  \big| X=x \right] \nonumber
\\
&=  \int \ell(y\,|\,  x;\theta,\eta) \,  [D^m_\eta u(z;\theta,\eta,\mu)] dy
\nonumber \\
& \quad +
 \sum_{\emptyset \neq S \subseteq \{1,\ldots,|m|\}}    (-1)^{|S|}   
     \sum_{T \subsetneq S}   
     \sum_{R \subseteq -S}
     \int   [D^{m_{R \cup T}}_\eta \, \ell(y\,|\,  x;\theta,\eta)] [ D^{m_{(-S \setminus R) \cup (S \setminus T)}}_\eta \, u(z;\theta,\eta,\mu)]  dy,
  \label{BruteForceDmu}   
\end{align}
where we write $-S$ for the set $\{1,\ldots,|m|\} \setminus S$. All the terms on the right hand side
of the last display equation are of the form
$ \int   [D^{m_A}_\eta \, \ell(y\,|\,  x;\theta,\eta)] [ D^{m_{-A}}_\eta \, u(z;\theta,\eta,\mu)]  dy$,
for some $A \subseteq \{1,\ldots,|m|\}$,
and we can therefore write
\begin{align}
  \mathbb{E}_{\theta,\eta}\left[  D^m_\eta u_q^*(Z;\theta,\eta,\mu)  \big| X=x \right] 
&=    \sum_{A \subseteq \{1,\ldots,|m|\}}  \,  \kappa_{A} \,  \int   [D^{m_A}_\eta \, \ell(y\,|\,  x;\theta,\eta)] [ D^{m_{-A}}_\eta \, u(z;\theta,\eta,\mu)]  dy ,
   \label{RewriteAsKappaSum}
\end{align}
with
\begin{align*}
\kappa_A = \mathbbm{1}\{A=\emptyset\} + \sum_{\emptyset \neq S \subseteq \{1,\ldots,|m|\}}    (-1)^{|S|}   
     \sum_{T \subsetneq S}   
     \sum_{R \subseteq -S}
     \mathbbm{1}\{R \cup T = A\}  .
\end{align*}
Here, the indicator $ \mathbbm{1}\{A=\emptyset\} $ accounts for the term $\int \ell(y\,|\,  x;\theta,\eta)  [D^m_\eta u(z;\theta,\eta,\mu)] dy$
in \eqref{BruteForceDmu},
while the second term in $\kappa_A$ counts the contributions from the triple sum.
Our goal is to show that $\kappa_A=0$ for all $A \subseteq \{1,\ldots,|m|\}$. We analyze two cases separately:

\begin{itemize}
\item
\textbf{Case 1:}
For \( A = \emptyset \), we note that \( R \cup T = \emptyset \) implies $R = T =  \emptyset$. Thus, the indicator \( \mathbbm{1}\{R \cup T = \emptyset\} \) is non-zero only when \( R = \emptyset \) and \( T = \emptyset \), implying that
\begin{align*}
\kappa_\emptyset 
 = 1 + \sum_{\emptyset \neq S \subseteq \{1, \ldots, |m|\}} (-1)^{|S|}
=   \sum_{S \subseteq \{1, \ldots, |m|\}} (-1)^{|S|}
= 0,
\end{align*}
where in the second step we used that
$1=(-1)^{|\emptyset|}$ to include that term into the sum over
$S$, and the final step is the alternating sum result
that is also used in the proof of Lemma~\ref{lemma:helpful}.

\item
\textbf{Case 2:}
Next, consider $A \neq \emptyset$. 
For given $A$ and $S$, we have
\begin{align}
\sum_{T \subsetneq S} \sum_{R \subseteq -S} \mathbbm{1}\{R \cup T = A\} =
\begin{cases}
1 & \text{if } S \not\subseteq A , \\
0 & \text{if }  S \subseteq A. \\
\end{cases}
   \label{MoreUsefulCombinatorics}
\end{align}
If $S \not\subseteq A$ (i.e.\ not $S \subseteq A$), then \eqref{MoreUsefulCombinatorics}
holds because a solution to
the conditions $R \cup T = A$, $T \subsetneq S$, $R \subseteq -S$ 
exists and is uniquely given by
$T = A \cap S$ and $R = A \cap (-S)$. 
Uniqueness of the pair $(T,R)$ implies that 
 $\sum_{T \subsetneq S}   
     \sum_{R \subseteq -S}
     \mathbbm{1}\{R \cup T = A\}=1$ in that case.
However, if $S \subseteq A$ then no solution for the pair 
$(T,R)$ exists (because $T = A \cap S$ implies $T=S$ in that case,
which contradicts the condition $T \subsetneq S$),
implying that the expression in \eqref{MoreUsefulCombinatorics} is indeed zero then.
Using  \eqref{MoreUsefulCombinatorics} we now find that
\begin{align*}
\kappa_A &= \sum_{\substack{\emptyset \neq S \subseteq \{1, \ldots, |m|\} \\ S \not\subseteq A}} (-1)^{|S|}
\\
&= \sum_{\emptyset \neq S \subseteq \{1, \ldots, |m|\}} (-1)^{|S|} - \sum_{\emptyset \neq S \subseteq A} (-1)^{|S|}
\\
&= \left[\sum_{S \subseteq \{1, \ldots, |m|\}} (-1)^{|S|} - (-1)^{|\emptyset|}\right] - \left[\sum_{S \subseteq A} (-1)^{|S|} - (-1)^{|\emptyset|}\right]
\\
&= [0 - 1] - [0 - 1] = 0.
\end{align*}

\end{itemize}

We have thus shown that $\kappa_A=0$ for all $A \subseteq \{1,\ldots,|m|\}$.
By \eqref{RewriteAsKappaSum} we thus have
$ \mathbb{E}_{\theta,\eta}\left[  D^m_\eta u_q^*(Z;\theta,\eta,\mu)  \big| X=x \right]=0$, which can also be written as
$$
\mathbb{E}_{\theta,\eta} \left[  \nabla^q_\eta \, u_q^*(Z;\theta,\eta,\mu)  \, \big| \, X=x \right] = 0 .
$$
Plugging in the true parameter values $\theta_0$ and $\eta_0$ we thus find
$$
\mathbb{E} \left[  \nabla^q_\eta \, u_q^*(Z;\theta_0,\eta_0,\mu)  \, \big| \, X=x \right] = 0 ,
$$
and by the law of iterated expectations also
$$
\mathbb{E} \left[  \nabla^q_\eta \, u_q^*(Z;\theta_0,\eta_0,\mu)  \right] = 0 .
$$
Remarkably, this result holds for any value of $\mu$.

\end{proof}

\subsection{\bf Proof of Lemma~\ref{lemma:Expansion}}

Let $\tau:=(\theta,\mu)$. 
We write $u_{k,i}(\tau, \eta_i)$ for
 $u_{k}(Z_i;   \theta, \eta_i ,  \mu)$, the $k$th component
 of the ${\rm dim}(u)$-vector $u(Z_i;   \theta, \eta_i ,  \mu)$.
Furthermore, compared to the statement of the lemma
we drop all subscripts $n$ in the following derivations.
In particular, for ${\mathbb K}_{q,n,i}$ we simply write
${\mathbb K}_{q,i}$.
By a mean-value expansion of $ \widehat \eta_i$ around $\eta_{i0}$ we obtain
\begin{align*}
   \frac 1 n \sum_{i=1}^N  u_{k,i}(Z_i;  \widehat \theta, \widehat \eta_i, \widehat \mu )
 &=
  \frac 1 n \sum_{i=1}^N  u_{k,i}( \widehat \tau, \widehat \eta_i )   
  \\
   &=  \frac 1 n \sum_{i=1}^N  u_{k,i}(  \widehat \tau, \eta_{i0} )   
\\ &\quad
  +
 \frac 1 n \sum_{i=1}^N \sum_{m \in {\mathbb K}_{q,i}} \frac{1}{m!} \left[ D^{m}_{\eta_i} \, u_{k,i}(  \widehat \tau, \eta_{i0} ) \right]   \left(\widehat \eta_i - \eta_{i0}\right)^{m} 
\\ &\quad
 + \frac 1 n \sum_{i=1}^N \sum_{m \in {\mathbb K}_{q+1,i} \setminus {\mathbb K}_{q,i}} \frac{1}{m!} \left[D^{m}_{\eta_i} \, u_{k,i}(   \widehat \tau, \widetilde \eta_i )\right]   \left(\widehat \eta_i - \eta_{i0}\right)^{m} ,
\end{align*}
where $m!=\prod_r (m_r!)$, and
 $\widetilde \eta_i$ is some value between $\eta_{i0}$ and $ \widehat \eta_i $.
Next, we perform a mean-value expansion in $\widehat \tau$ around $\tau_0$ to obtain
\begin{align*}
   \frac 1 n \sum_{i=1}^N  u_{k,i}(Z_i;  \widehat \theta, \widehat \eta_i, \widehat \mu )
  &=
  \frac 1 n \sum_{i=1}^N  u_{k,i}(  \tau_0, \eta_{i0} )   
\displaybreak[1] \\ &\quad
  +
\underbrace{ \frac 1 n \sum_{i=1}^N \left[  \frac{\partial } {\partial \tau}  u_{k,i}(   \tau_0, \eta_{i0} )  \right]^\top  \left(\widehat \tau - \tau_0\right)
}_{= \left[G_\mu (\widehat \mu - \mu_0) + G_\theta (\widehat \theta - \theta_0)\right]_k + o_P(\left\| \widehat \tau - \tau_0\right\| )} 
\displaybreak[1] \\ &\quad
  + 
  \underbrace{
   \frac 1 2  \left(\widehat \tau - \tau_0\right)^\top 
  \left\{
 \frac 1 {n} \sum_{i=1}^N  \left[  \frac{\partial^2} {\partial \tau \partial \tau^\top}  u_{k,i}(  \widetilde \tau, \eta_{i0} )  \right] 
 \right\}  \left(\widehat \tau - \tau_0\right) 
 }_{=:B_{1,k}}
\displaybreak[1] \\ &\quad
  +
\underbrace{ \frac 1 n \sum_{i=1}^N \sum_{m \in {\mathbb K}_{q,i}} \frac{1}{m!} \left[ D^{m}_{\eta_i} \, u_{k,i}(  \tau_0, \eta_{i0} ) \right]  \left(\widehat \eta_i - \eta_{i0}\right)^{m} 
}_{= n^{-1/2} \, R_{n,k} , \; \text{the $k$'th component of $R_n$ defined in the lemma.}}
\displaybreak[1] \\ &\quad
  +
\underbrace{  \frac 1 n \sum_{i=1}^N \sum_{m \in {\mathbb K}_{q,i} } \frac{1}{m!} \left[ D^{m}_{\eta_i}  \, \frac{\partial} {\partial \tau} u_{k,i}(   \overline \tau, \eta_{i0} )   \right]^\top
 \left(\widehat \tau - \tau_0\right)  \left(\widehat \eta_i - \eta_{i0}\right)^{m} 
 }_{=: B_{2,k}}
\displaybreak[1] \\ &\quad
+ \underbrace{    \frac 1 n \sum_{i=1}^N \sum_{m \in {\mathbb K}_{q+1,i} \setminus {\mathbb K}_{q,i} } \frac{1}{m!} \left[D^{m}_{\eta_i} \, u_{k,i}(   \widehat \tau, \widetilde \eta_i )\right]   \left(\widehat \eta_i - \eta_{i0}\right)^{m} 
 }_{=: B_{3,k}}
 ,
\end{align*}
where $\widetilde \tau$ and $\overline \tau$ are values between $\widehat \tau$ and $\tau_0$.
Denote the dimensions of the parameters $\theta$ and $\mu$ by $d_\theta$ and $d_\mu$, respectively.
Our assumptions guarantee that
\begin{align*}
    | B_{1,k} | &\leq \frac {(d_\theta+d_\mu)^2} 2 \left\| \widehat \tau - \tau_0 \right\|^2  \frac{1}{n} \sum_{i=1}^N n_i  C(Z_i)   \\
   &\leq \frac {(d_\theta+d_\mu)^2} 2 \left\| \widehat \tau - \tau_0 \right\|^2 \underbrace{ \left( \frac{1}{n} \sum_{i=1}^N n_i \left[ C(Z_i) \right]^2 \right)^{1/2}}_{=O_P(1)}\underbrace{ \left( \frac{1}{n} \sum_{i=1}^N  n_i \right)^{1/2}}_{=O(1)} = O_P \left( \left\| \widehat \tau - \tau_0 \right\|^2 \right),
   \end{align*}
\begin{align*}     | B_{2,k} | &\leq
     (d_\theta+d_\mu) \,
      \left\| \widehat \tau - \tau_0\right\| \,
  \frac 1 n \sum_{i=1}^N  n_iC(Z_i)   \sum_{m \in {\mathbb K}_{q,i} }   \frac{1}{m!}  \big| \left( \widehat \eta_i - \eta_{i0} \right)^m \big|       \\
     &\leq  (d_\theta+d_\mu) \,
      \left\| \widehat \tau - \tau_0\right\| \,
\underbrace{  \left( \frac{1}{n} \sum_{i=1}^N  n_i\left[ C(Z_i) \right]^2 \right)^{1/2} }_{=O_P(1)}
  \underbrace{
   \left( \frac 1 n \sum_{i=1}^N n_i
   \left[   \sum_{m \in {\mathbb K}_{q,i} }   \frac{1}{m!}  \big| \left( \widehat \eta_i - \eta_{i0} \right)^m \big| \right]^2
   \right)^{1/2}  }_{=o_P(1)}
   \\
     &= o_P\left(  \left\| \widehat \tau - \tau_0\right\| \right) ,
 \end{align*}
 \begin{align*}
      | B_{3,k} | &\leq  \frac 1 n \sum_{i=1}^N n_i C(Z_i)    \left\| \widehat \eta_i - \eta_{i0} \right\|^{q+1}
    \underbrace{  \sum_{ m \in {\mathbb K}_{q+1,i}  \setminus {\mathbb K}_{q,i}} \frac{1}{m!} }_{=O(1)}
  \\
   &=   O(1) 
   \underbrace{  \left( \frac{1}{n} \sum_{i=1}^N  n_i\left[ C(Z_i) \right]^2 \right)^{1/2} }_{=O_P(1)}
     \underbrace{ 
   \left( \frac 1 n \sum_{i=1}^N n_i
 \left\| \widehat \eta_i - \eta_{i0} \right\|^{2(q+1)}
   \right)^{1/2}  }_{= o_P(n^{-1/2})} =  o_P(n^{-1/2}).
\end{align*}
Here, in addition to our assumption we also used the Cauchy-Schwarz inequality.
We have thus shown that
\begin{align*}
   \frac 1 n \sum_{i=1}^N  u(Z_i;  \widehat \theta, \widehat \eta_i, \widehat \mu )
  &=
  G_\mu (\widehat \mu - \mu_0) + G_\theta (\widehat \theta - \theta_0) 
  + \frac 1 n \sum_{i=1}^N  u(Z_i;  \theta_0, \eta_{i0}, \mu_0 )      +  n^{-1/2} R_{n} 
 \\ &\quad
  + 
 O_P \left( \left\| \widehat \tau - \tau_0 \right\|^2 \right)
  +o_P\left(  \left\| \widehat \tau - \tau_0\right\| \right)
 + o_P(n^{-1/2})
 .
\end{align*}
Using our assumptions on the convergence of $\widehat \mu $ and $\widehat \theta$ we thus have
\begin{align}
   \frac 1 {\sqrt{n}} \sum_{i=1}^N  u(Z_i;  \widehat \theta, \widehat \eta_i, \widehat \mu )
  &=
   G_\mu \left[ \sqrt{n} (\widehat \mu - \mu_0) \right]  +  o_P\!\left( 1 + \left\| \sqrt{n}(\widehat \mu - \mu_0)\right\| \right)
 \nonumber \\ &\quad
   +
     \frac 1 {\sqrt{n}} \sum_{i=1}^N \left[ u(Z_i;   \theta_0, \eta_{i0} ,  \mu_0) +  G_{\theta} \, \psi_i \right]
                   + R_n  + o_P(1).
                \label{uExpansions}   
\end{align}
By Assumption~\ref{ass:asymptotic1}(i)
we have 
$$\left[\frac 1 n \sum_{i=1}^N  
    \frac{\partial u^\top (Z_i;  \widehat \theta,\widehat \eta_{i},\widehat\mu  )} {\partial \mu}
     \right] W \left[\frac 1 {\sqrt{n}} \sum_{i=1}^N  u(Z_i;  \widehat \theta,\widehat \eta_{i},\widehat\mu  ) \right]  =  o_P(1).
$$     
By using Assumption~\ref{ass:asymptotic1}(iv) and (vii) we thus have
$G_\mu^\top W \left[\frac 1 {\sqrt{n}} \sum_{i=1}^N  u(Z_i;  \widehat \theta,\widehat \eta_{i},\widehat\mu  ) \right]  =  o_P(1)$, and plugging the approximation in \eqref{uExpansions} then gives
\begin{align*}
   o_P(1) &=
  G_\mu^\top W \left[ \frac 1 {\sqrt{n}} \sum_{i=1}^N  u(Z_i;  \widehat \theta, \widehat \eta_i, \widehat \mu )
  \right]
  \\
  &=
  \left( G_\mu^\top W G_\mu \right) \left[ \sqrt{n} (\widehat \mu - \mu_0) \right]  +   o_P\!\left( 1 + \left\| \sqrt{n}(\widehat \mu - \mu_0)\right\| \right)
 \nonumber \\ &\quad
   +
   G_\mu^\top W
   \left\{ \frac 1 {\sqrt{n}} \sum_{i=1}^N \left[ u(Z_i;   \theta_0, \eta_{i0} ,  \mu_0) +  G_{\theta} \, \psi_i \right]  +   R_n  \right\}
                   + o_P(1).
\end{align*}
Since $G_\mu^\top W G_\mu $ is full rank, absorbing the $o_P(\|\sqrt{n}(\widehat\mu-\mu_0)\|)$ term into
$G_\mu^\top W G_\mu$ and solving for $\sqrt{n} (\widehat \mu - \mu_0)$
gives the statement of the lemma.

\subsection{Proof of Theorem~\ref{th:Asymptotic}} 

We again drop all subscripts $n$ in the  derivations. Let $\xi_{i} = u(Z_i; \theta_0, \eta_{i0}, \mu_{0}) +G_{\theta} \psi_{i}$. From Lemma~\ref{lemma:Expansion}, we have
\begin{align*}
    \sqrt{n}(\widehat{\mu} - \mu_{0}) &= -(G_{\mu}^{\top}WG_{\mu})^{-1}G_{\mu}^{\top}W\left\{\frac{1}{\sqrt{n}}\sum_{i=1}^N \xi_{i} + R_n\right\} + o_P(1).
\end{align*}
We will show that $R_n = o_P(1)$ under our assumptions. First, by Assumption~\ref{ass:asymptotic2}(i), the moment function is Neyman-orthogonal to order $q$, which implies
\begin{align*}
    \mathbb{E} \left[D^m_{\eta_{i}}u(Z_i; \theta_0, \eta_{i0}, \mu_{0})\right] = 0
\end{align*}
for all $m \in {\mathbb{K}}_{q,i}$. Therefore, $R_n$ is a sum of mean-zero terms.
Next, by Assumption~\ref{ass:asymptotic2}(ii), (iii), the summands of $R_n$ are uncorrelated across $i$ (since, conditional on $(\widehat\eta_{1},\ldots,\widehat\eta_{N})$, they are independent with zero conditional means),
so that $\mathbb{E}[\|R_n\|^2]$ is $\frac 1 n \sum_{i=1}^N$ of their second moments. By the second-moment bound in
Assumption~\ref{ass:asymptotic1}(iv) and Assumption~\ref{ass:asymptotic2}(ii), the latter are bounded by
a constant times $n_i \widetilde C_{n,i}\,$ $c_{n,i}$, where $c_{n,i} := \big( \mathbb{E}\big\|\widehat \eta_{i}-\eta_{i0}\big\|^{2(q+1)} \big)^{\frac{1}{q+1}}$. Here we used that, by Lyapunov's inequality,
$\mathbb{E}\big[ \big( (\widehat \eta_{i}-\eta_{i0})^{m} \big)^{2} \big] \leq \mathbb{E}\big\|\widehat \eta_{i}-\eta_{i0}\big\|^{2|m|} \leq c_{n,i}^{\,|m|}$,
and that $c_{n,i}^{\,|m|} \leq c_{n,i}$ for all $1 \leq |m| \leq q$ once $n$ is sufficiently large, because Assumption~\ref{ass:asymptotic1}(v) together with $n_i \geq 1$ implies $\max_i c_{n,i} = o(1)$. Splitting
$n_i = \sqrt{n_i}\sqrt{n_i}$ and applying the Cauchy-Schwarz inequality as above,
\begin{align*}
\frac 1 n \sum_{i=1}^N n_i \widetilde C_{n,i}\, c_{n,i}
\leq \underbrace{ \left( \frac 1 n \sum_{i=1}^N n_i \widetilde C_{n,i}^2 \right)^{1/2} }_{=O(1)}
\underbrace{ \left( \frac 1 n \sum_{i=1}^N n_i \, c_{n,i}^{2} \right)^{1/2} }_{=o(n^{-1/(q+1)})} ,
\end{align*}
where the second factor follows from Jensen's inequality applied to the
concave map $x \mapsto x^{2/(q+1)}$ (note that $c_{n,i}^{2} = (c_{n,i}^{\,q+1})^{2/(q+1)}$) together with $\frac 1 n \sum_{i=1}^N n_i = 1$, and
Assumption~\ref{ass:asymptotic1}(v). Hence $\mathbb{E}[\| R_n\|^2 ] = o(1)$.
By Chebyshev's inequality we thus have $R_n = o_P(1)$.
Thus, we have
\begin{align*}
    \sqrt{n}(\widehat{\mu} - \mu_{0}) &= -(G_{\mu}^{\top}WG_{\mu})^{-1}G_{\mu}^{\top}W\left\{\frac{1}{\sqrt{n}}\sum_{i=1}^N \xi_{i}\right\} + o_P(1).
\end{align*}
By Assumption~\ref{ass:asymptotic2}(ii), (iii), the terms $\xi_{i}$ are independent across $i$. Furthermore, by Assumption~\ref{ass:asymptotic2}(iv), they satisfy Lindeberg's condition and have a well-defined variance limit $V_{\xi}$. Therefore, by the Lindeberg-Feller Central Limit Theorem:
\begin{align*}
    \frac{1}{\sqrt{n}}\sum_{i=1}^N \xi_{i} \overset{d}{\to} \mathcal{N}(0, V_{\xi}).
\end{align*}
The conclusion follows by the continuous mapping theorem, giving us
\begin{align*}
    \sqrt{n}(\widehat{\mu} - \mu_{0}) \overset{d}{\to} \mathcal{N}\left(0, (G_{\mu}^{\top}WG_{\mu})^{-1}G_{\mu}^{\top}WV_{\xi}WG_{\mu}(G_{\mu}^{\top}WG_{\mu})^{-1}\right).
\end{align*}

\clearpage

\begin{center}
	{ {\large SUPPLEMENT TO: \\``A NEYMAN-ORTHOGONALIZATION APPROACH TO\\THE INCIDENTAL PARAMETER PROBLEM\\IN LIKELIHOOD MODELS''} }
\end{center}

\begin{center}
St\'ephane Bonhomme, Koen Jochmmans, and Martin Weidner
\end{center}

\renewcommand{\theequation}{\thesection.\arabic{equation}}  \setcounter{equation}{0}

\section{Proof of Lemma~\ref{lemma:helpful}\label{App_prooflemma}}

By taking derivatives of $\int \ell(y|x;\theta,\eta) dy = 1$ with respect to $\eta$, we obtain
$$
\int [\nabla^q_\eta \ell(y|x;\theta,\eta)] dy = 0,
$$
which can also be written as $\mathbb{E}_{\theta,\eta}[w_q(Z;\theta,\eta)|X=x] = 0$. Since 
$\varSigma_{w_qw_q}(X;\theta,\eta)$ does not 
depend on $Y$ we also have
$\mathbb{E}_{\theta,\eta}[\widetilde{w}_q(Z;\theta,\eta)|X=x] = 0$.
Using this and the definition of $\widetilde{w}_q$
we obtain
\begin{align}
\mathbb{I}_{k_q} &= \varSigma_{w_qw_q}(x;\theta,\eta)^{-1}\varSigma_{w_qw_q}(x;\theta,\eta) 
\nonumber \\
&= \varSigma_{w_qw_q}(x;\theta,\eta)^{-1}\mathbb{E}_{\theta,\eta}[w_q(Z;\theta,\eta)w_q(Z;\theta,\eta)^\top|X=x] 
\nonumber \\
&= \mathbb{E}_{\theta,\eta}[\widetilde{w}_q(Z;\theta,\eta)w_q(Z;\theta,\eta)^\top|X=x] \nonumber \\
&= \int \widetilde{w}_q(z;\theta,\eta)[\nabla^q_\eta \ell(y|x;\theta,\eta)]^\top dy.
\label{DerivationTildeWlemma}
\end{align}
According to its definition in Subsection~\ref{subsec:DefHigherOrder}, the elements of
the $k_q$-vector operator $\nabla^q_\eta$
are given by
$$
  D^m_\eta = \frac{\partial^p}{\partial \eta_{m_1}\cdots \partial \eta_{m_p}} ,
$$
and are uniquely labeled by vectors of 
integers $m = (m_1,\ldots, m_p)$ of length
$p \in \{1,\ldots,q\}$ in the following set
$$
   \mathcal{C}_q = \bigcup_{p \in \{1,\ldots,q\}} \,
\lbrace m=(m_1,\ldots, m_p): 1\leq m_1\leq \cdots \leq m_p \leq d_\eta \rbrace.   
$$
Analogously, we now introduce the notation
$\widetilde{w}_q^{\,m}(z;\theta,\eta)$, $m \in \mathcal{C}_q$,
to uniquely denote the elements of the 
vector $\widetilde{w}_q(z;\theta,\eta)$, which is also
a vector of length $k_q = |  \mathcal{C}_q |$.
With that notation, the result in display~\eqref{DerivationTildeWlemma} can equivalently be
written as
\begin{align}
    \forall r,m \in \mathcal{C}_q: \qquad
    \int \widetilde{w}_q^{\,r}(z;\theta,\eta) \,
         [D^{m}_\eta \ell(y|x;\theta,\eta)] \,
         dy
    &= \mathbbm{1}\{r=m\} .
    \label{DerivationTildeWlemma2}
\end{align}
Since $\mathbb{E}_{\theta,\eta}[\widetilde{w}_q(Z;\theta,\eta)|X=x] = 0$, we also have
\begin{align*}
    \int \widetilde{w}_q^{\,r}(z;\theta,\eta) \, \ell(y|x;\theta,\eta) \,
         dy
    &= 0 .
\end{align*}
For the empty vector $()$ of length zero we have $D^{()}_\eta \ell(y|x;\theta,\eta) = \ell(y|x;\theta,\eta)$. Using this notation
we can combine the result in the last two displays to find that for all $r    \in \mathcal{C}_q$ and all $v \in \mathcal{C}_q \cup \{()\}$
we have
\begin{align*}
    \int \widetilde{w}_q^{\,r}(z;\theta,\eta) \,
         [D^{v}_\eta \ell(y|x;\theta,\eta)] \,
         dy
    &= \mathbbm{1}\{r=v\} .
\end{align*}
For $k \in \{1,2,3,\ldots\}$, vector $t=(t_1,\ldots,t_k) \in \mathcal{C}_q$, and a subset $S \subseteq \{1,\ldots,k\}$, let $t_S$ denote the vector formed by keeping only the indices in $S$, and let $t_{-S} = t_{\{1,\ldots,k\}\setminus S}$ be the vector of the remaining elements.
Then, by applying $D^t_\eta$ to the last display and using the product rule for differentiation we obtain
\begin{align} 
 \sum_{S \subseteq \{1,\ldots,k\}} \int [D_\eta^{t_{S}} \widetilde{w}_q^{\,r}(z;\theta,\eta)] \, [D_\eta^{t_{-S}} D_\eta^v \ell(y|x;\theta,\eta)] \, dy = 0 .
     \label{ApplyProductRule}
 \end{align}
Of course, we have $D_\eta^{t} D_\eta^{v}  = D^{(t,v)}_\eta$, and instead of distinguishing between $t$ and $v$ we can also just write $m$ for $(t,v)$
combined. The last display equation then implies that  for any nonempty subset $T  \subseteq \{1,2,\ldots,|m|\}$ we have 
(just set $t=m_T$ and $v=m_{-T}$ in the last display result---in doing so, it was important that above we allowed for $v$  to be the empty vector):
 \begin{align*} 
 \sum_{S \subseteq T} \int [D_\eta^{m_{S}} \widetilde{w}_q^{\,r}(z;\theta,\eta)] \, [D_\eta^{m_{-S}}   \ell(y|x;\theta,\eta)] \, dy = 0 .
 \end{align*}
 Now, consider the following linear combination of the result in the last display,
 \begin{align} 
   \sum_{\substack{T \subseteq \{1,2,\ldots,|m|\}
   \\ T \neq \emptyset}} (-1)^{|T|} \sum_{S \subseteq T} \int [D_\eta^{m_{S}} \widetilde{w}_q^{\,r}(z;\theta,\eta)] \, [D_\eta^{m_{-S}}   \ell(y|x;\theta,\eta)] \, dy = 0 .
     \label{LinearCombinationOverT}
 \end{align}
 For a fixed  $S  \subseteq \{1,2,\ldots,|m|\}$,  the total coefficient of the term 
$\int [D_\eta^{m_{S}} \widetilde{w}_q^{\,r}] \, [D_\eta^{m_{-S}} \ell] \, dy$ in this linear combination is given by
\begin{align}
\sum_{\substack{T \, : \, S \subseteq T \subseteq \{1,\ldots,|m|\} \\ T \neq \emptyset}} (-1)^{|T|} 
   &= 
\begin{cases} 
-1 & \text{if } S = \emptyset, \\ 
(-1)^{|m|} & \text{if } S = \{1,\ldots,|m|\}, \\ 
0 & \text{otherwise}.
\end{cases}
   \label{LemmaCombinatoricsResult}
\end{align}
To see that the result in the last display holds, notice first that
for $S = \emptyset$ the sum is simply 
$$
   \sum_{ T \, : \, \emptyset \neq T} \subseteq \{1,\ldots,|m|\} (-1)^{|T|}
   =  -1 +
  \underbrace{ \sum_{T \subseteq \{1, \ldots, |m|\}} (-1)^{|T|}
  }_{=0} = -1,
$$
where $\sum_{T \subseteq \{1, \ldots, |m|\}} (-1)^{|T|}=\sum_{r=0}^{|m|} {|m| \choose r} (-1)^r=(1-1)^{|m|}=0$ is a
classic alternating sum result, which holds for all $|m| \geq 1$. 
Next, for $S = \{1,\ldots,|m|\}$, the left hand side of \eqref{LemmaCombinatoricsResult} only sums over one element,
$T=\{1,\ldots,|m|\}$, and we thus get $(-1)^{|T|} = (-1)^{|m|}$ for the sum.
Finally, if $S \neq \emptyset$ and $S \neq \{1,\ldots,|m|\}$, then
the left hand side of \eqref{LemmaCombinatoricsResult} can be written
as
\begin{align*}
\sum_{T \, : \, S \subseteq T \subseteq \{1,\ldots,|m|\}} (-1)^{|T|} 
&= \sum_{R \subseteq -S} (-1)^{|S| + |R|}   
= (-1)^{|S|} \underbrace{ \sum_{R \subseteq -S} (-1)^{|R|}  }_{=0}
 =0
\end{align*}
where we replaced the sum over $T$ by a sum over $R$ 
such that $T = S \cup R$, with $-S = \{1, \ldots, |m|\} \setminus S$,
and in the final step we used the alternating sum result again.

Using \eqref{LemmaCombinatoricsResult},  our linear combination in \eqref{LinearCombinationOverT} equals
\begin{align*}
- \int [D_\eta^{m} \widetilde{w}_q^{\,r}(z;\theta,\eta)] \, \ell(y|x;\theta,\eta) \, dy 
+ (-1)^{|m|}   \int\widetilde{w}_q^{\,r}(z;\theta,\eta)  [D_\eta^{m}  \ell(y|x;\theta,\eta)] \, dy = 0 .
\end{align*}
Together with \eqref{DerivationTildeWlemma2} we thus find
\begin{align*}
\int [D_\eta^{m} \widetilde{w}_q^{\,r}(z;\theta,\eta)] \, \ell(y|x;\theta,\eta) \, dy
  &=  (-1)^{|m|} \,  \mathbbm{1}\{r=m\}  ,
\end{align*}
which in vector notation can be written as \eqref{lemma:helpful:result}.

\renewcommand{\theequation}{\thesection.\arabic{equation}}  \setcounter{equation}{0}

\section{Order $q$ of orthogonality\label{App_q}}

\subsection{A diagnostic for $q$}

Consider a moment function $u(Z_i;\theta,\eta_i,\mu)$, and denote its $q$-orthogonal counterparts based on Theorem \ref{theo_neyman} as $u_q^*(Z_i;\theta,\eta_i,\mu)$, for $q=1,2,...$. Given some order $q$ we propose the following diagnostic:
	$$S_q=\left\|\frac{1}{\sqrt{n}}\sum_{i=1}^N\left[u_q^*(Z_i;\widehat\theta, \widehat \eta_i,\widehat \mu)-u_{q+1}^*(Z_i;\widehat\theta,\widehat \eta_i,\widehat \mu)\right]\right\|^2_{\widehat V^{-1}},$$
where $\widehat\theta$ and $\widehat\mu$ are based on $(q+1)$-orthogonal functions, and $\widehat V$ is a consistent estimator of the variance 
	$$V=\underset{n\rightarrow \infty}{\mbox{lim}}\, \mbox{Var}\left[\frac{1}{\sqrt{n}}\sum_{i=1}^N(u_q^*(Z_i;\widehat\theta, \eta_i,\widehat \mu)-u_{q+1}^*(Z_i; \widehat\theta,\eta_i,\widehat \mu))\right].$$

As a special case, suppose that one wishes to estimate $\theta$, and $u(Z_i;\theta,\eta_i)$ does not depend on $\mu$. Then, since $\widehat\theta$ is based on $u_{q+1}^*$ that is uncorrelated with $u_{q}^*-u_{q+1}^*$, a consistent estimator of $V$ is
	$$\widehat{V}=\frac{1}{n}\sum_{i=1}^N (u_q^*(Z_i; \widehat\theta,\widehat\eta_i)-u_{q+1}^*(Z_i; \widehat\theta,\widehat\eta_i))(u_q^*(Z_i; \widehat\theta,\widehat\eta_i)-u_{q+1}^*(Z_i; \widehat\theta,\widehat\eta_i))^\top.$$
    More generally, a consistent estimator of $V$ is available using standard Taylor expansions and Lemma \ref{lemma:Expansion}. We provide an example in the case of the Neyman-Scott model in the next subsection.

    Under the conditions of Theorem \ref{th:Asymptotic}, which in particular guarantee that the estimates $\widehat\eta_i$ converge at a fast enough rate, $S_q$ converges in distribution to a $\chi^2_r$ random variable, where $r=\dim u$. In practice, we thus suggest comparing $S_q$ to the quantiles of the $\chi^2_r$ distribution.

\subsection{Illustration in the Neyman-Scott model}

Consider the estimation of $\mu = \nicefrac{1}{N} \sum_{i=1}^N h(\eta_i)$ in the Neyman-Scott model introduced in Subsection \ref{subsec:Setup}. Here we assume that $T$ -- the number of periods used in estimation -- is fixed as $N$ tends to infinity, and consequently we use $\sqrt{N}$ in order to scale all cross-sectional average quantities. Suppose first that $\sigma$ is known. By (\ref{eq_ustar_NS}) we have 
\begin{align*}&u_{q}^*(Y_i; \sigma^2, \eta_i,\mu)-u_{q+1}^*(Y_i; \sigma^2, \eta_i,\mu)\\&=- \sigma^{q+1}T^{-\frac{q+1}{2}}\frac{1}{(q+1)!}\nabla_{\eta}^{(q+1)}h(\eta_i)H_{q+1}\left(\frac{\sum_{j=1}^T(Y_{ij}-\eta_{i})}{ \sqrt{T}\sigma}\right),\end{align*}
and
$$V=\underset{N\rightarrow \infty}{\mbox{lim}}\,\sigma^{2q+2}T^{-q-1}\frac{1}{(q+1)!}\frac{1}{N}\sum_{i=1}^N[\nabla_{\eta}^{(q+1)}h(\eta_i)]^2.$$
Hence, again treating $\sigma$ as known, the diagnostic is
\begin{equation}S_q=\frac{\left[\sum_{i=1}^N\nabla_{\eta}^{(q+1)}h(\widehat\eta_i)H_{q+1}\left(\frac{\sum_{j=1}^T(Y_{ij}-\widehat\eta_{i})}{ \sqrt{T}\sigma}\right)\right]^2}{(q+1)!\sum_{i=1}^N[\nabla_{\eta}^{(q+1)}h(\widehat \eta_i)]^2}.\label{eq_Sq}\end{equation}

With $\sigma$ unknown, we have, using $\widehat{\sigma}$ in (\ref{eq_sig2_hat}),
\begin{align*}&u_{q}^*(Y_i; \widehat\sigma^2, \eta_i,\mu)-u_{q+1}^*(Y_i; \widehat\sigma^2, \eta_i,\mu)\\&=- \widehat\sigma^{q+1}T^{-\frac{q+1}{2}}\frac{1}{(q+1)!}\nabla_{\eta}^{(q+1)}h(\eta_i)H_{q+1}\left(\frac{\sum_{j=1}^T(Y_{ij}-\eta_{i})}{ \sqrt{T}\widehat\sigma}\right)\\
&=-\sigma^{q+1}T^{-\frac{q+1}{2}}\frac{1}{(q+1)!}\nabla_{\eta}^{(q+1)}h(\eta_i)H_{q+1}\left(\frac{\sum_{j=1}^T(Y_{ij}-\eta_{i})}{ \sqrt{T}\sigma}\right)\\&\quad+\frac{1}{2T}\widetilde\sigma^{q-1}T^{-\frac{q-1}{2}}\frac{1}{(q-1)!}\nabla_{\eta}^{(q+1)}h(\eta_i)H_{q-1}\left(\frac{\sum_{j=1}^T(Y_{ij}-\eta_{i})}{ \sqrt{T}\widetilde\sigma}\right)(\widehat\sigma^2-\sigma^2),\end{align*}
for $\widetilde\sigma$ between $\sigma$ and $\widehat\sigma$, where we have used the expression for the derivatives of Hermite polynomials.

Hence, for $q\geq 2$ we have
\begin{align*}&\frac{1}{\sqrt{N}}\sum_{i=1}^N\left(u_{q}^*(Y_i; \widehat\sigma^2, \eta_i,\mu)-u_{q+1}^*(Y_i; \widehat\sigma^2, \eta_i,\mu)\right)\\&=-\frac{1}{\sqrt{N}}\sum_{i=1}^N\sigma^{q+1}T^{-\frac{q+1}{2}}\frac{1}{(q+1)!}\nabla_{\eta}^{(q+1)}h(\eta_i)H_{q+1}\left(\frac{\sum_{j=1}^T(Y_{ij}-\eta_{i})}{ \sqrt{T}\sigma}\right)+o_P(1),\end{align*}
and we rely on the diagnostic $S_q$ given by (\ref{eq_Sq}) even though $\sigma$ is estimated. Now, for $q=1$ we have
\begin{align*}&\frac{1}{\sqrt{N}}\sum_{i=1}^N\left(u_{1}^*(Y_i; \widehat\sigma^2, \eta_i,\mu)-u_{2}^*(Y_i; \widehat\sigma^2, \eta_i,\mu)\right)\\&=-\frac{1}{\sqrt{N}}\sum_{i=1}^N\sigma^{2}T^{-1}\frac{1}{2}\nabla_{\eta}^{(2)}h(\eta_i)\left(\left(\frac{\sum_{j=1}^T(Y_{ij}-\eta_{i})}{ \sqrt{T}\sigma}\right)^2-1\right)
\\&\quad+\frac{1}{\sqrt{N}}\sum_{i=1}^N\frac{1}{2T}\nabla_{\eta}^{(2)}h(\eta_i)\left(\frac{1}{N(T-1)} \sum_{i'=1}^N \sum_{j=1}^T (Y_{i'j}-\overline{Y}_{i'})^2-\sigma^2\right)+o_P(1),\end{align*}
where we have used the expressions of $H_2$ and $\widehat\sigma^2$. That is,
\begin{align*}&\frac{1}{\sqrt{N}}\sum_{i=1}^N\left(u_{1}^*(Y_i; \widehat\sigma^2, \eta_i,\mu)-u_{2}^*(Y_i; \widehat\sigma^2, \eta_i,\mu)\right)\\&=\frac{1}{\sqrt{N}}\sum_{i=1}^N\frac{1}{2T}\nabla_{\eta}^{(2)}h(\eta_i)\left[\frac{1}{N(T-1)} \sum_{i'=1}^N \sum_{j=1}^T (Y_{i'j}-\overline{Y}_{i'})^2-\left(\frac{\sum_{j=1}^T(Y_{ij}-\eta_{i})}{ \sqrt{T}}\right)^2\right]+o_P(1).\end{align*}
In this case, $$V=\underset{N\rightarrow \infty}{\mbox{lim}}\,\sigma^{4}T^{-2}\frac{1}{2}\left[\frac{1}{N}\sum_{i=1}^N[\nabla_{\eta}^{(2)}h(\eta_i)]^2+\frac{1}{T-1}\left(\frac{1}{N}\sum_{i=1}^N\nabla_{\eta}^{(2)}h(\eta_i)\right)^2\right],$$ and we rely on 
$$S_1=\frac{\left\{\sum_{i=1}^N\nabla_{\eta}^{(2)}h(\widehat\eta_i)\left[\left(\frac{\sum_{j=1}^T(Y_{ij}-\widehat\eta_{i})}{ \sqrt{T}\widehat{\sigma}}\right)^2-\frac{1}{N(T-1)} \sum_{i'=1}^N \sum_{j=1}^T \left(\frac{Y_{i'j}-\overline{Y}_{i'}}{\widehat\sigma}\right)^2\right]\right\}^2}{2\left[\sum_{i=1}^N[\nabla_{\eta}^{(2)}h(\widehat\eta_i)]^2+\frac{1}{N(T-1)}\left(\sum_{i=1}^N\nabla_{\eta}^{(2)}h(\widehat\eta_i)\right)^2\right]}.$$

\begin{remark}{(sample splitting)}
When implementing the diagnostic in this example, we use a small number $T$ of time periods in estimation. Keeping $T$ fixed -- while at the same time having $\widehat{\eta}_i\overset{p}{\rightarrow}\eta_i$ -- is important for the theory in order to guarantee that the impact of the estimation error in $\widehat{\eta}_i$ does not affect the asymptotic distribution of $S_q$. In practice, one can cross-fit by taking multiple subsets of $T$ time periods (along with different $\widehat{\eta}_i$'s) and average over them to reduce variability. 
\end{remark}

\renewcommand{\theequation}{\thesection.\arabic{equation}}  \setcounter{equation}{0}

\section{Implementation in nonlinear regression\label{App_sec_implement}}

\subsection{Expression for $M$}

Following \citet{constantine1996multivariate}, consider a multivariate function
$$  h\left(x_1, \ldots, x_d\right)=f\left[g^{(1)}\left(x_1, \ldots, x_d\right), \ldots, g^{(m)}\left(x_1, \ldots, x_d\right)\right],
$$
$h_\nu=D_{\mathbf{x}}^\nu h\left(\mathbf{x}^0\right), f_{\boldsymbol{\lambda}}=D_{\mathbf{y}}^{\boldsymbol{\lambda}} f\left(\mathbf{y}^0\right), g_{\boldsymbol{\mu}}^{(i)}=D_{\mathbf{x}}^\mu g^{(i)}\left(\mathbf{x}^0\right), \mathbf{g}_{\boldsymbol{\mu}}=\left(g_{\boldsymbol{\mu}}^{(1)}, \ldots, g_{\boldsymbol{\mu}}^{(m)}\right)$.
The Fa\`a di Bruno formula is (Theorem 2.1 in \citealp{constantine1996multivariate}):
$$
h_{\boldsymbol{\nu}}=\sum_{1 \leq|\boldsymbol{\lambda}| \leq n} f_{\boldsymbol{\lambda}}\, \underset{\text{elements of }M}{\underbrace{\sum_{s=1}^n \sum_{p_s(\boldsymbol{\nu}, \boldsymbol{\lambda})}(\boldsymbol{\nu} !) \prod_{j=1}^s \frac{\left[\mathbf{g}_{\ell_j}\right]^{\mathbf{k}_j}}{\left(\mathbf{k}_{j} !\right)\left[\boldsymbol{\ell}_{j} !\right]^{\left|\mathbf{k}_j\right|}}}},
$$
where $n=|\boldsymbol{\nu}|$, and
$$
\begin{gathered}
	p_s(\boldsymbol{\nu}, \boldsymbol{\lambda})=\bigg\{\left(\mathbf{k}_1, \ldots, \mathbf{k}_s ; \boldsymbol{\ell}_1, \ldots, \boldsymbol{\ell}_s\right):\left|\mathbf{k}_i\right|>0, \\
	\mathbf{0} \prec \boldsymbol{\ell}_1 \prec \cdots \prec \boldsymbol{\ell}_s, \sum_{i=1}^s \mathbf{k}_i=\boldsymbol{\lambda} \text { and } \sum_{i=1}^s\left|\mathbf{k}_i\right| \boldsymbol{\ell}_i=\boldsymbol{\nu}\bigg\} .
\end{gathered}
$$

\subsection{Useful properties of the normal distribution}

We first consider the univariate normal case.
\begin{lemma}
	\label{lemma:univariate}
	For $m \in \mathbb{R}$ and $\sigma \in (0,\infty)$, let $ Y \sim {\cal N}(m,\sigma^2)$, with corresponding likelihood function
	$$
	\ell(y\,|\,m,\sigma)  = \frac {1} { \sqrt{2\pi \sigma^2} }  \exp\left( - \frac{(y-m)^2} {2 \, \sigma^2} \right)  
	.
	$$
	Let   $j,k \in \{0,1,2,\ldots \}$, and define
	\begin{align*}
		\kappa_{jk}
		&:= \mathbb{E}_{m,\sigma} \left[ \frac 1 { \ell(Y\,|\,m,\sigma) }  \, \frac{\partial^{j} \ell(Y\,|\,m,\sigma)} {(\partial m)^{j} } \,
		\frac 1 { \ell(Y\,|\,m,\sigma) }  \, \frac{\partial^{k} \ell(Y\,|\,m,\sigma)} {(\partial m)^{k} } \,
		\right] ,
		\\
		\rho_{j}
		&:= \mathbb{E}_{m,\sigma} \left[ \frac 1 { \ell(Y\,|\,m,\sigma) }  \, \frac{\partial^{j} \ell(Y\,|\,m,\sigma)} {(\partial m)^{j} } \,
		\frac{\partial  \log \ell(Y\,|\,m,\sigma)} { \partial \sigma  }  \,
		\right] .
	\end{align*}      
	Then,  
	$$
	\kappa_{jk}  =    \mathbbm{1} \big\{ j=k \big\} \;  \frac{j!} {\sigma^{2j}}  \;  ,
	\qquad   \qquad
	\rho_j  =   \mathbbm{1} \big\{ j=2 \big\}   \frac{2 } {\sigma^{3}}  \; .
	$$
\end{lemma}
Let $ \phi(y)   = \frac {1} { \sqrt{2\pi} }  \exp\left( -   y^2 / 2  \right)$ and
$ \phi^{(j)}(y)  = \frac{d^j \phi(y) } {dy^j } $. Hermite polynomials are defined by
$h_j(y) = (-1)^j [  \phi(y) ]^{-1} \, \phi^{(j)}(y)  $.  The proof of Lemma~\ref{lemma:univariate} is given in Subsection \ref{sec_prooflemma7}.
It crucially relies on the following orthogonality property of Hermite polynomials:
\begin{align}
	\int_{-\infty}^\infty  h_j(y) \, h_k(y) \, \phi(y)  \,   dy =  \mathbbm{1} \big\{ j=k \big\} \;  j!  \;.
	\label{HermiteOrthogonality}
\end{align}
The result in Lemma~\ref{lemma:univariate} is sufficient for our purposes, but more general results can be derived.\footnote{
	More generally, for   $k_1,k_2 \in \{0,1,2,\ldots \}$ and $j_1,j_2 \in \{0,1\}$, let
	\begin{align*}
		\kappa_{k_1,k_2,j_1,j_2}
		&:= \mathbb{E}_{m,\sigma} \left[ \frac 1 { \ell(Y\,|\,m,\sigma) }  \, \frac{\partial^{k_1+j_1} \ell(Y\,|\,m,\sigma)} {(\partial m)^{k_1} (\partial \sigma)^{j_1} } \,
		\frac 1 { \ell(Y\,|\,m,\sigma) }   \,      \frac{\partial^{k_2+j_2} \ell(Y\,|\,m,\sigma)} {(\partial m)^{k_2} (\partial \sigma)^{j_2} }
		\right]
		\\
		&=  \int_{-\infty}^\infty  \frac 1 { \ell(y\,|\,m,\sigma) }  \, \frac{\partial^{k_1+j_1} \ell(y\,|\,m,\sigma)} {(\partial m)^{k_1} (\partial \sigma)^{j_1} } \,
		\frac{\partial^{k_2+j_2} \ell(y\,|\,m,\sigma)} {(\partial m)^{k_2} (\partial \sigma)^{j_2} } dy.
	\end{align*}
	One then finds
	\begin{align*}
		\kappa_{k_1,k_2,j_1,j_2} =  \mathbbm{1} \big\{k_1+2 j_1=k_2 + 2 j_2 \big\}  \;  (k_1+2 j_1)! \; \sigma^{-[2 (k_1+2 j_1) - j_1 - j_2]}.
	\end{align*}
}

Next, we consider a vector of independent normal variables with heteroscedastic means and variances.

\begin{lemma}
	\label{lemma:multivariate}
	Let $d \in \{1,2,3,\ldots\}$.
	For $m \in \mathbb{R}^d$ and $\sigma \in (0,\infty)^d$, let
	$\Sigma(\sigma)$ be the $d \times d$ diagonal matrix with diagonal entries $\sigma_i^2$, and let
	$ Y \sim {\cal N}(m,\Sigma(\sigma))$. The corresponding likelihood function reads
	\begin{align*}
		\ell(y\,|\,m,\sigma)  &=  \prod_{i=1}^d \ell(y_i\,|\,m_i,\sigma_i) ,
		&
		\ell(y_i\,|\,m_i,\sigma_i) &=
		\left[ \frac {1} { \sqrt{2\pi \sigma_i^2} }  \exp\left( - \frac{(y_i-m_i)^2} {2 \, \sigma_i^2} \right)   \right].
	\end{align*}
	Let $j,k \in \{0,1,2,\ldots \}^d$,   $j^* = \sum_{i=1}^d j_i$, $k^* = \sum_{i=1}^d k_i$, and define
	\begin{align*}
		\kappa(j,k)
		&:= \mathbb{E}_{m,\sigma} \left[ \frac 1 { \ell(Y\,|\,m,\sigma) }  \, \frac{\partial^{j^*} \ell(Y\,|\,m,\sigma)} {\prod_{i=1}^d  (\partial m_i)^{j_i} } \,
		\frac 1 { \ell(Y\,|\,m,\sigma) }  \,   \frac{\partial^{k^*} \ell(Y\,|\,m,\sigma)} {\prod_{i=1}^d  (\partial m_i)^{k_i} } \,
		\right] ,
		\\
		\rho(j,i')
		&:= \mathbb{E}_{m,\sigma} \left[ \frac 1 { \ell(Y\,|\,m,\sigma) }  \, \frac{\partial^{j^*} \ell(Y\,|\,m,\sigma)} {\prod_{i=1}^d  (\partial m_i)^{j_i} }\,
		\frac{\partial  \log \ell(Y\,|\,m,\sigma)} { \partial \sigma_{i'}  }  \,
		\right] ,
	\end{align*}
	where $i' \in \{1,\ldots,d\}$ in the last line.      
	Then,  
	$$
	\kappa(j,k)  =    \mathbbm{1} \big\{ j=k \big\} \;  \prod_{i=1}^d  \frac{j_i!} {\sigma_i^{2j_i}}  \;  ,
	\qquad    
	\rho(j,i)  = \left\{ \begin{array}{ll}  \displaystyle   \frac{2 } {\sigma_i^{3}}       & \text{if $j_i = 2$, and all other entries of $j$ are zero,}
		\\[20pt]
		0 & \text{otherwise.}
	\end{array} \right.
	$$
\end{lemma}

Lemma~\ref{lemma:multivariate} is an immediate corollary of Lemma~\ref{lemma:univariate}. Using the independence
of the components of $Y$ we find
$$
\frac{\partial^{j^*} \ell(y\,|\,m,\sigma)} {\prod_{i=1}^d  (\partial m_i)^{j_i}} 
=  \prod_{i=1}^d \frac{\partial^{j_i} \ell(y_i\,|\,m_i,\sigma_i)} {  (\partial m_i)^{j_i} } ,
$$
and
$$
\kappa(j,k)  =   \prod_{i=1}^d \, \kappa_{j_i,k_i} .
$$
Plugging in the result for $\kappa_{jk}$ in Lemma~\ref{lemma:univariate} then gives the result for $    \kappa(j,k) $
in Lemma~\ref{lemma:multivariate}. Analogously for $ \rho(j,i')$.

\subsection{Nonlinear regression with normal errors}

{Consider the model}
\begin{align*}
	Y_i &= m(X_i;\theta,\eta) + \sigma(X_i;\theta)  \, U_i ,
	&
	U_i \,|\, X &\sim iid{\cal N}(0,1) ,
	&
	i &= 1,\ldots,d,
\end{align*}
where $m(\cdot;\cdot,\cdot)$ and  $\sigma(\cdot;\cdot)$ are known functions, and $\theta$ and $\eta$ are unknown parameters.
Ignore $\theta$ for the moment and write 
\begin{align*}
	Y_i &= m_i(\eta) + \sigma_i  \, U_i .
\end{align*}
Let $m=(m_1,\ldots,m_d)$ and $\sigma=(\sigma_1,\ldots,\sigma_d)$.
The likelihood for  $y=(y_1,\ldots,y_d)$ is then given by
$$
\ell(y\,|\,\eta) = \ell(y\, | \,m(\eta),\sigma) ,
$$
where $\ell(y\,|\,m,\sigma)$ is given in Lemma~\ref{lemma:multivariate}.
Let $\nabla_{\eta}^p$ be the vector operator that collects all unique derivatives with respect to $\eta$ up to order $p$. Let $\nabla^{p}_m$ be the vector operator that collects all unique derivatives with respect to $m$ up to order $p$. Then, there exists a matrix valued function $M(\eta)$, which only depends on $\eta$ and on the function $m(\eta)$, such that
\begin{align}
	\nabla_{\eta}^p \,  \ell(y\,|\,\eta) &=   M(\eta) \, \nabla_{m}^p \,  \ell(y\,|\,m(\eta),\sigma) .
	\label{FaaDiBruno}
\end{align}

We want to calculate
\begin{align*}
	\mathbb{E}_{\eta} \left[ \frac {\nabla^{p}_{\eta}  \ell(Y\,|\,\eta) } { \ell(Y\,|\,\eta) }     \,
	\frac {\nabla^{p }_{\eta}  \ell(Y\,|\,\eta)^{\top}} { \ell(Y\,|\,\eta) }  
	\right].
\end{align*}
Lemma~\ref{lemma:multivariate}  gives us explicit expressions for all the components of 
\begin{align*}
	\mathbb{E}_{m,\sigma} \left[ \frac {\nabla^{p}_m  \ell(Y\,|\,m,\sigma) } { \ell(Y\,|\,m,\sigma) }     \,
	\frac {\nabla^{p }_m  \ell(Y\,|\,m,\sigma)^{\top}} { \ell(Y\,|\,m,\sigma) }  
	\right].
\end{align*}
Using \eqref{FaaDiBruno} we have
\begin{align*}
	&\mathbb{E}_{\eta} \left[ \frac {\nabla^{p}_{\eta}  \ell(Y\,|\,\eta) } { \ell(Y\,|\,\eta) }     \,
	\frac {\nabla^{p }_{\eta}  \ell(Y\,|\,\eta)^{\top}} { \ell(Y\,|\,\eta) }  
	\right]\\
	&=   M(\eta) \,    \mathbb{E}_{m(\eta),\sigma} \left[ \frac {\nabla^{p}_m  \ell(Y\,|\,m(\eta),\sigma) } { \ell(Y\,|\,m(\eta),\sigma) }     \,
	\frac {\nabla^{p }_m  \ell(Y\,|\,m(\eta),\sigma)^{\top}} { \ell(Y\,|\,m(\eta),\sigma) }  
	\right]  \,  M(\eta)^{\top} . 
\end{align*}
Thus, by combining Lemma~\ref{lemma:multivariate} with the multidimensional Faà di Bruno's formula we get explicit expressions for all the matrices we need.

\subsection{Proof of Lemma~\ref{lemma:univariate}\label{sec_prooflemma7}}

	We already introduced $ \phi(y)   = \frac {1} { \sqrt{2\pi} }  \exp\left( -   y^2 / 2  \right)$ and
	$ \phi^{(j)}(y)  = \frac{d^j \phi(y) } {dy^j } $ above.
	Let  $j,k \in \{0,1,2,3,\ldots\}$.
	The well-known  orthogonality property  of Hermite polynomials in \eqref{HermiteOrthogonality} can be rewritten as
	\begin{align}
		\int_{-\infty}^\infty  \frac {  \phi^{(j)}(y)  \,    \phi^{(k)}(y) } { \phi(y)  }  \,   dy =  \mathbbm{1} \big\{ j=k \big\} \;  j!  \;.
		\label{HermiteOrthogonality2}  
	\end{align}
	Another well-known  property  of Hermite polynomials is the recurrence relation
	$ h_{j+1}(y) = y  h_j(y) -  \frac{d} {dy} h_j(y) $. Using this, it is easy to show that
	for $j>k$ we have
	\begin{align}
		\int_{-\infty}^\infty  \frac {  y \, \phi^{(j)}(y)  \,    \phi^{(k)}(y) } { \phi(y)  }  \,   dy =  - \; \mathbbm{1} \big\{ j=k+1 \big\} \;   j!  \;.
		\label{HermiteOrthogonality3}  
	\end{align}
	Next, we have
	\begin{align*}
		\ell(y\,|\,m,\sigma) &= \frac 1 {\sigma} \, \phi\left(  \frac{y-m} \sigma  \right)  ,
		&
		\frac{\partial^{j} \ell(y\,|\,m,\sigma)} {(\partial m)^{j}  } 
		&=  \frac{ (-1)^{j} } {\sigma^{j+1}}  \;  \phi^{(j)}\left(  \frac{y-m} \sigma  \right)  .
	\end{align*}   
	Using this we obtain
	\begin{align*}
		\kappa_{jk}
		&:= \mathbb{E}_{m,\sigma} \left[ \frac 1 { \ell(Y\,|\,m,\sigma) }  \, \frac{\partial^{j} \ell(Y\,|\,m,\sigma)} {(\partial m)^{j} } \,
		\frac 1 { \ell(Y\,|\,m,\sigma) }  \, \frac{\partial^{k} \ell(Y\,|\,m,\sigma)} {(\partial m)^{k} } \,
		\right]
		\\
		&=  \int_{-\infty}^\infty  \frac 1 { \ell(y\,|\,m,\sigma) }  \, \frac{\partial^{j} \ell(y\,|\,m,\sigma)} {(\partial m)^{j}   } \,
		\frac{\partial^{k} \ell(y\,|\,m,\sigma)} {(\partial m)^{k}   } \, dy
		\\
		&=    \frac{ (-1)^{j+k} } {\sigma^{j+k+1}}  \, \int_{-\infty}^\infty  \frac 1 {  \phi\left(  \frac{y-m} \sigma  \right)  }  \, \phi^{(j)}\left(  \frac{y-m} \sigma  \right) \,
		\phi^{(k)}\left(  \frac{y-m} \sigma  \right) \, dy
		\\
		&=    \frac{ (-1)^{j+k} } {\sigma^{j+k}}  \,   \int_{-\infty}^\infty  \frac {  \phi^{(j)}(y)  \,    \phi^{(k)}(y) } { \phi(y)  }  \,   dy
		\\
		&=          \mathbbm{1} \big\{ j=k \big\} \;  \frac{j!} {\sigma^{2j}}   \; ,
	\end{align*}
	where  the second to last step employs a change of variables in the integral ($\frac{y-m} \sigma \mapsto y$), and the last step uses \eqref{HermiteOrthogonality2}.
	Similarly, for 
	\begin{align*}
		\frac{\partial  \ell(y\,|\,m,\sigma)} { \partial \sigma  } 
		&=   -  \frac{1} {\sigma^2}  \, \phi\left(  \frac{y-m} \sigma  \right)
		-  \left(\frac{y-m} {\sigma^3}\right) \, \phi^{(1)}\left(  \frac{y-m} \sigma  \right) ,
	\end{align*}   
	one finds
	\begin{align*}
		\rho_{j}
		&:= \mathbb{E}_{m,\sigma} \left[ \frac 1 { \ell(Y\,|\,m,\sigma) }  \, \frac{\partial^{j} \ell(Y\,|\,m,\sigma)} {(\partial m)^{j} } \,
		\frac{\partial  \log \ell(Y\,|\,m,\sigma)} { \partial \sigma  }  \,
		\right]
		\\
		&=  \int_{-\infty}^\infty  \frac 1 { \ell(y\,|\,m,\sigma) }  \, \frac{\partial^{j} \ell(y\,|\,m,\sigma)} {(\partial m)^{j}   } \,
		\frac{\partial  \ell(y\,|\,m,\sigma)} { \partial \sigma  }  \, dy
		\\
		&=    \frac{ (-1)^{1+j} } {\sigma^{2+j}}  \, \int_{-\infty}^\infty  \frac 1 {  \phi\left(  \frac{y-m} \sigma  \right)  }  \, \phi^{(j)}\left(  \frac{y-m} \sigma  \right) \,
		\left[ 
		\phi\left(  \frac{y-m} \sigma  \right)
		+ \left(\frac{y-m} {\sigma}\right) \, \phi^{(1)}\left(  \frac{y-m} \sigma  \right)
		\right] \, dy
		\\
		&=    \frac{ (-1)^{1+j} } {\sigma^{1+j}}  \, \int_{-\infty}^\infty  \frac 1 {  \phi\left( y  \right)  }  \, \phi^{(j)}\left( y  \right) \,
		\left[ 
		\phi\left( y  \right)
		+  y \, \phi^{(1)}\left( y  \right)
		\right] \, dy
		\\
		&=     \mathbbm{1} \big\{ j=2 \big\}   \frac{ (-1) } {\sigma^{3}}  \, \int_{-\infty}^\infty  \frac { y \, \phi^{(1)}\left( y  \right)  \, \phi^{(2)}\left( y  \right)} {  \phi\left( y  \right)  }   \, dy
		\\
		&=     \mathbbm{1} \big\{ j=2 \big\}   \frac{2 } {\sigma^{3}},
	\end{align*}
	where we again employed the same change of variables in the integration and also used \eqref{HermiteOrthogonality2} and \eqref{HermiteOrthogonality3}.

\renewcommand{\theequation}{\thesection.\arabic{equation}}  \setcounter{equation}{0}

\section{Monte Carlo simulation\label{AppC}}

In this section of the appendix we report on the results of a Monte Carlo experiment. We specify a CES model of team production with log-normal errors, where we take the network structure (i.e., the set ${\cal{K}}$ in (\ref{eq_teams})) as given from the empirical data. We fix the true value of the substitution parameter to $\gamma_0=1$, the team size parameter to $\beta_{0}=1$, the log-error variance in teams of size 2 to $\sigma_{0}^2(2)=1$, and the variance in teams of size 1 to $\sigma_{0}^2(1)=1$. This data generating process is designed to approximate what we found on the empirical data. 

We report results based on 300 simulations. In  each simulated sample, we estimate the parameters using plug-in method-of-moments and the Neyman-orthogonalized method-of-moments estimates of degree $q=1$ to $q=6$. As we did in our empirical study, we compute sample-split preliminary estimates of the author fixed-effects based on all their sole-authored publications except for one, selected at random. However, in the simulation exercise we do not cross-fit the estimators, and simply choose a random selection of sole-authored publications for each author in each Monte Carlo run. 

In Tables \ref{tab_simu_1} and \ref{tab_simu_2} we show the median, mean, 2.5\% quantile, and 97.5\% quantile of each estimate across simulations. Starting with the substitution parameter $\gamma$, we see that the plug-in estimator is severely biased, with median and mean biases of -50\% (expressed in proportion of the true value). For this parameter, all Neyman-orthogonalized estimators are substantially less biased, with a median bias below 3\%, and below 1\% for $q\geq 4$. However, in some replications the orthogonalized estimates tend to have large values, which is reflected in a somewhat larger mean bias, ranging from 6\% at $q=2$ down to 1.5\% at $q=5$ and $q=6$, and quantile bands that are not symmetric around the true value.  

Turning next to the team size parameter $\beta$, we see that both the plug-in and first-order Neyman-orthogonalized estimators are biased, with a median and mean bias of 4\%--6\%. All orthogonalized estimates of order $q\geq 2$ are virtually unbiased, both for the mean and the median. Moreover, in this case the quantile bands are symmetric around the true parameter value.

  \begin{table}
	\caption{Monte Carlo simulation\label{tab_simu_1}}
	\begin{center}
		\begin{tabular}{|c||cccc|}\hline\hline
			\multicolumn{5}{|c|}{Substitution ${\gamma}$ (true value=1)}\\\hline\hline
			& Median & Mean & 2.5\% &  97.5\%\\\hline
			Plug-in &  0.5054 & 0.5089 & 0.4150 & 0.5966\\
			$q=1$ &  0.9723 & 0.9864 & 0.7523 & 1.3259 \\ 
			$q=2$ &  1.0310 & 1.0625 & 0.7787 & 1.5571\\
			$q=3$ &  1.0100 & 1.0408 & 0.7455 & 1.5795 \\ 
			$q=4$ &  0.9964 & 1.0227 & 0.7364 & 1.5556\\ 
			$q=5$ &  0.9898 & 1.0161 & 0.7363 & 1.4904 \\ 
			$q=6$ &  0.9897 & 1.0154 & 0.7362 & 1.4904 \\\hline\hline
			\multicolumn{5}{|c|}{Team size ${\beta}$ (true value=1)}\\\hline\hline
			& Median & Mean & 2.5\% &  97.5\%\\\hline
			Plug-in &  1.0578 & 1.0595 & 1.0277 & 1.1003  \\ 
			$q=1$ &  1.0459 & 1.0466 & 0.9971 & 1.1038  \\ 
			$q=2$ &  1.0029 & 1.0012 & 0.9299 & 1.0605 \\ 
			$q=3$ &  1.0007 & 0.9979 & 0.9249 & 1.0577     \\ 
			$q=4$ &  1.0013 & 0.9980 & 0.9226 & 1.0590  \\ 
			$q=5$ &  1.0007 & 0.9983 & 0.9203 & 1.0566 \\ 
			$q=6$ &  1.0007 & 0.9983 & 0.9214 & 1.0565 \\\hline\hline
		\end{tabular}
	\end{center}
	
	\raggedright\textit{{\footnotesize Notes: 300 simulations.}}
\end{table}

Shifting attention to the variance in teams of size 2, $\sigma^2(2)$, we see that both the plug-in and first-order Neyman-orthogonalized estimators are severely biased, with a median and mean bias of 16\%--18\%. All orthogonalized estimates of order $q\geq 2$ are virtually unbiased, both for the mean and the median, and the quantile bands are centered around the true parameter value.

Lastly, turning to the variance in teams of size 1, $\sigma^2(1)$, the plug-in estimator exhibits a large bias of 33\%. First-order orthogonalization only decreases the bias slightly, to 27\%. In contrast, the Neyman-orthogonalized estimators continue to show good performance. In particular, when $q\geq 4$ the estimates are virtually unbiased, and the quantile bands are symmetric around the true value.

\begin{table}
 	\caption{Monte Carlo simulation (continued)\label{tab_simu_2}}
 	\begin{center}
 		\begin{tabular}{|c||cccc|}\hline\hline
 			\multicolumn{5}{|c|}{Variance ${\sigma^2(2)}$ (true value=1)}\\\hline\hline
 			& Median & Mean & 2.5\% &  97.5\%\\\hline
 			Plug-in & 1.1623 & 1.1618 & 1.1284 & 1.1944\\
 			$q=1$ & 1.1824 & 1.1825 & 1.1456 & 1.2228 \\	
 			$q=2$ & 0.9964 & 0.9969 & 0.9615 & 1.0334 \\ 
 			$q=3$ & 1.0001 & 1.0009 & 0.9665 & 1.0366 \\ 
 			$q=4$ & 0.9995 & 1.0006 & 0.9637 & 1.0373 \\ 
 			$q=5$ & 0.9993 & 1.0000 & 0.9639 & 1.0365 \\ 
 			$q=6$ & 0.9982 & 0.9996 & 0.9637 & 1.0374  \\\hline\hline
 			\multicolumn{5}{|c|}{Variance ${\sigma^2(1)}$ (true value=1)}\\\hline\hline
 			& Median & Mean & 2.5\% &  97.5\%\\\hline
 			Plug-in & 1.3361 & 1.3368 & 1.2840 & 1.3856\\
 			$q=1$ & 1.2732 & 1.2723 & 1.2140 & 1.3299\\
 			$q=2$ & 1.0158 & 1.0178 & 0.9478 & 1.0925\\
 			$q=3$ & 1.0087 & 1.0119 & 0.9435 & 1.0842\\
 			$q=4$ & 0.9996 & 1.0033 & 0.9335 & 1.0761\\
 			$q=5$ & 0.9980 & 1.0002 & 0.9321 & 1.0731\\
 			$q=6$ & 0.9976 & 0.9992 & 0.9272 & 1.0718\\\hline\hline
 		\end{tabular}
 	\end{center}
 	
 	\raggedright\textit{{\footnotesize Notes: 300 simulations.}}
 \end{table}

 \renewcommand{\theequation}{\thesection.\arabic{equation}}  \setcounter{equation}{0}

 \section{Restrictions independent of individual effects\label{AppD}}

\begin{figure}
 	\caption{Comparing two estimates of $\beta_0$\label{fig_comparebeta}}
 	\begin{center}
 		\includegraphics[width=100mm, height=60mm]{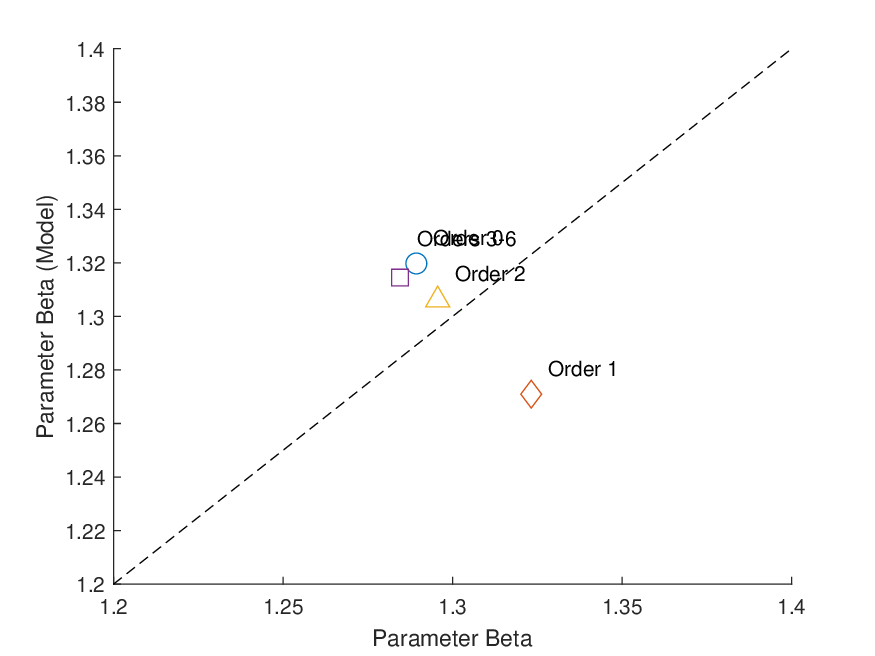}
 	\end{center}
 	\par
 	\raggedright\textit{{\footnotesize Notes: Estimate of $\beta_0$ on the x-axis, model-based estimate of $\beta_0$ based on the right-hand side of (\ref{eq_2beta}) on the y-axis. Each point corresponds to an order of orthogonalization. }}
 \end{figure}

 Model (\ref{eq_teams}) implies restrictions on parameters $\gamma_0,\beta_0,\sigma_0^2(1),\sigma_0^2(2)$ that do not depend on the author-specific effects $\eta_{i0}$.\footnote{The analysis in this section was inspired by discussions with Bo Honor\'e.} As an example, the model implies the following alternative expression for the team size parameter $\beta_0$: 
 \begin{equation}
 	\beta_0=\left(\frac{\mathbb{E}[Y_j^{\gamma_0}\,|\, s_j=2]}{\mathbb{E}[Y_j^{\gamma_0}\,|\, s_j=1]}\right)^{\frac{1}{\gamma_0}}\exp\left(\frac{1}{2}\gamma_0[\sigma_0^2(1)-\sigma_0^2(2)]\right),\label{eq_2beta}
 \end{equation}
where the expectation given $s_j=1$ is taken over the co-authors' own sole-authored articles, and (\ref{eq_2beta}) does not involve the fixed-effects $\eta_{i0}$. Note that, if $\gamma_0=0$ and log-output is additive in worker inputs, then $\log\beta_0$ is simply the difference between average log-outputs in teams of size 2 and 1, respectively. As a check, in Figure \ref{fig_comparebeta} we report estimates of the left-hand side of (\ref{eq_2beta}), against the estimates of $\beta_0$ shown in Table \ref{tab_appli}, for various orders of orthogonalization. We see that the estimates of the two sides of (\ref{eq_2beta}) tend to agree with each other well irrespective of the orthogonalization order. 
 
Model (\ref{eq_teams}) also implies restrictions on $\gamma_0$ alone. To see this, let us write (\ref{eq_teams}), within teams of size 2 only, as
\begin{equation*}
	Y_j^{\gamma_0}=\frac{1}{2}\beta_0^{\gamma_0}\left(\eta_{k(j,1)0}^{\gamma_0}+\eta_{k(j,2)0}^{\gamma_0}\right)\varepsilon_j^{\gamma_0\sigma_0(2)},
\end{equation*}
which we write in vector form as
\begin{equation}
	Y(\gamma_0)=A\widetilde{\eta}_0+\widetilde{\varepsilon},\label{eq_Ygamma_vec}
\end{equation}
where $Y(\gamma_0)$ has elements $Y_j^{\gamma_0}$, $A$ is a matrix of zeros and ones, $$\widetilde{\eta}_{k0}=\frac{1}{2}\beta_0^{\gamma_0}\eta_{k0}^{\gamma_0}\exp\left(\frac{1}{2}\gamma_0^2\sigma_0^2(2)\right),$$ and $$\widetilde{\varepsilon}_j=\frac{1}{2}\beta_0^{\gamma_0}\left(\eta_{k(j,1)0}^{\gamma_0}+\eta_{k(j,2)0}^{\gamma_0}\right)\left[\varepsilon_j^{\gamma_0\sigma_0(2)}-\exp\left(\frac{1}{2}\gamma_0^2\sigma_0^2(2)\right)\right].$$ Since $\mathbb{E}[\widetilde{\varepsilon}_j\,|\, A]=0$, (\ref{eq_Ygamma_vec}) implies the conditional moment equalities 
\begin{equation}
	\mathbb{E}\left[(I-AA^{\dagger})Y(\gamma_0)\,|\,A\right]=0,\label{eq_quasidiff}
\end{equation}
which only depend on $\gamma_0$.

To use (\ref{eq_quasidiff}) for estimation, we rely on a set of instruments. For this purpose, we use interacted preliminary estimates $Z_j=\widehat{\eta}_{k(j,1)}\widehat{\eta}_{k(j,2)}$ for $k(j,1),k(j,2)$ the co-authors of $j$. Since we assume the preliminary estimates are constructed from an independent sample, we have
\begin{equation}
	\mathbb{E}\left[Z'(I-AA^{\dagger})Y(\gamma_0)\right]=0.\label{eq_quasidiff2}
\end{equation}
Note these restrictions remain valid when $\varepsilon_j$ are not Gaussian or not mutually independent, provided they are independent of $A$. We use GMM estimation based on (\ref{eq_quasidiff2}), that is,
\begin{equation}
	\widehat{\gamma}^{\rm GMM}=\underset{\gamma}{\rm {argmin}}\, |Z'(I-AA^{\dagger})Y(\gamma)|.\label{eq_GMMest}
\end{equation}
We implement this estimator in the same way we have implemented our Neyman-orthogonalized equations. Specifically, we construct preliminary estimates of author effects using all but one sole-authored paper for each author, where we select the held-out sole-authored paper at random.

Using the same Monte Carlo simulation design as in Section \ref{AppC} tends to give noisy estimates. For example, when the true value is $\gamma_0=1$, and $\sigma_0(1)=1/5$ and $\sigma_{0}(2)=1/5$, we obtain a mean GMM estimate of $1.0218$, a median estimate of $0.9725$, and a standard deviation of $0.2261$ across 300 simulations. Moreover, out of the 300 simulations, in 24 cases we are unable to find another minimum in (\ref{eq_GMMest}) other than $\gamma=0$ (which is always a minimum by construction). Note these findings correspond to a model with error variances that are \emph{25 times} smaller than the variances we used for our main simulation design in Section \ref{AppC}. This suggests this estimation approach, at least for this particular choice of instruments, is considerably less precise than our likelihood-based approach.  

Lastly, computing the GMM estimator on the empirical data, cross-fitting 100 times, we obtain $\widehat{\gamma}^{\rm GMM}=0.5807$. This is of a comparable magnitude to the estimates of $\gamma$ reported in Table \ref{tab_appli}, when using a sufficiently high order of orthogonalization. However, it is worth noting that, out of the 100 random splits of the sole-authored productions, in 5 cases we are unable to find another minimum in (\ref{eq_GMMest}) other than $\gamma=0$, again reflecting the instability of this method in our setting.

\renewcommand{\theequation}{\thesection.\arabic{equation}}  \setcounter{equation}{0}

\section{Estimated production function\label{App_fig}}

\begin{figure}[h!]
	\caption{Production function estimate\label{fig_prodf}}
	\begin{center}
		\includegraphics[width=100mm, height=60mm]{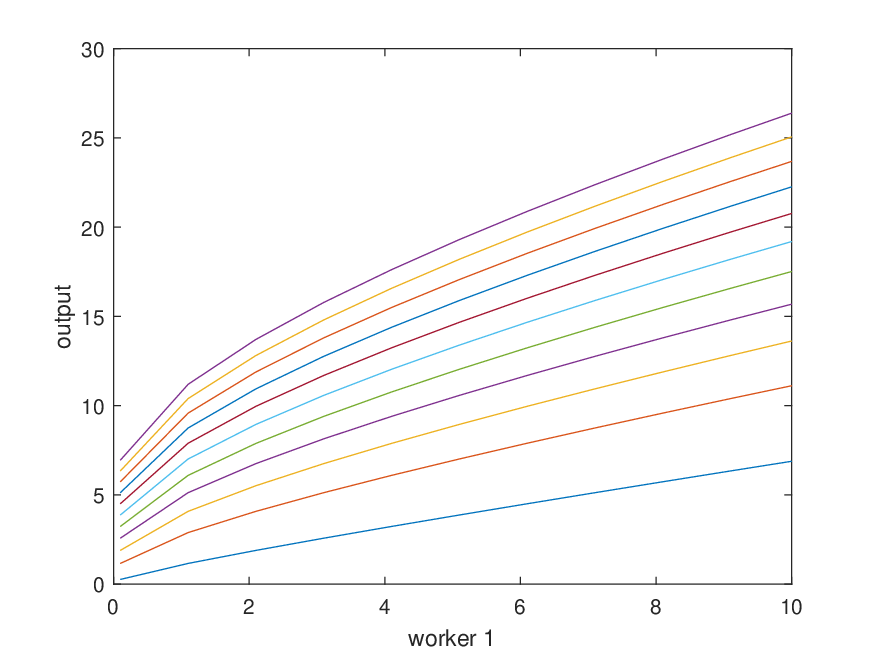}
	\end{center}
	\par
\raggedright	\textit{{\footnotesize Notes: Worker 1's type ${\eta_1}$ on the x-axis, average output $Y_j$ on the y-axis. Each curve corresponds to a different worker 2's type ${\eta_2}$.  Figure based on the point estimates for $q=6$ reported in Table \ref{tab_appli}.}}
\end{figure}

\end{document}